\documentclass[draft]{agujournal2019}%
\usepackage{url} 
\usepackage{lineno}
\usepackage{soul}
\usepackage{color}
\usepackage{xcolor}
\usepackage{soul}
\usepackage{caption}
\usepackage{subcaption}
\usepackage{comment} 
\usepackage{colortbl} 
\usepackage{listings}
\usepackage{blkarray} 
\usepackage{bbm}
\usepackage{bm}
\usepackage{array} 
\usepackage{setspace, verbatim, mathrsfs}
\usepackage{dsfont, bm}
\usepackage{empheq} 
\usepackage{amsmath}
\usepackage{ragged2e}

\draftfalse

\journalname{Water Resources Research}

\begin{document}

\justify
%
%


\title{Unsupervised anomaly detection in spatio-temporal stream network sensor data}

%
%





\authors{Edgar Santos-Fernandez\affil{1},
Jay M. Ver Hoef\affil{2},
Erin E. Peterson\affil{1},
James McGree\affil{1},
Cesar A. Villa \affil{6}, 
Catherine Leigh\affil{3},
Ryan Turner\affil{4,5},
Cameron Roberts\affil{4},
Kerrie Mengersen\affil{1}}

\affiliation{1}{School of Mathematical Sciences. Queensland University of Technology}
\affiliation{2}{Marine Mammal Laboratory. NOAA-NMFS Alaska Fisheries Science Center}
\affiliation{3}{Biosciences and Food Technology Discipline, School of Science. Royal Melbourne Institute of Technology (RMIT)}
\affiliation{4}{Water Quality \& Investigations.  Queensland Department of Environment and Science}
\affiliation{5}{School of Earth and Environmental Sciences, University of Queensland, Brisbane, Queensland, Australia}
\affiliation{6}{Science Information Services. Queensland Department of Environment and Science}

\correspondingauthor{Edgar Santos-Fernandez}{santosfe@qut.edu.au}




\begin{keypoints}
\item Technical anomalies in water quality data from in-situ sensors are common, posing challenges in their detection that often require substantial time and effort.
\item We present a flexible anomaly detection framework used to automatically detect multiple anomaly types using empirical model residuals.
\item 
To facilitate real-time monitoring, we have developed a Bayesian recursive stream-network model that effectively captures spatio-temporal dependencies within the data.
\item The timely identification of anomalies plays a crucial role in reducing uncertainties and biases, thereby enabling a more accurate understanding and management of water resources.
\end{keypoints}

\begin{abstract}
The use of in-situ digital sensors for water quality monitoring is becoming increasingly common worldwide. While these sensors provide near real-time data for science, the data are prone to technical anomalies that can undermine the trustworthiness of the data and the accuracy of statistical inferences, particularly in spatial and temporal analyses. Here we propose a framework for detecting anomalies in sensor data recorded in stream networks, which takes advantage of spatial and temporal autocorrelation to improve detection rates. The proposed framework involves the implementation of effective data imputation to handle missing data, alignment of time-series to address temporal disparities, and the identification of water quality events. We explore the effectiveness of a suite of state-of-the-art statistical methods including posterior predictive distributions, finite mixtures, and Hidden Markov Models (HMM). We showcase the practical implementation of automated anomaly detection in near-real time by employing a Bayesian recursive approach. This demonstration is conducted through a comprehensive simulation study and a practical application to a substantive case study situated in the Herbert River, located in Queensland, Australia, which flows into the Great Barrier Reef. We found that methods such as posterior predictive distributions and HMM produce the best performance in detecting multiple types of anomalies. Utilizing data from multiple sensors deployed relatively near one another enhances the ability to distinguish between water quality events and technical anomalies, thereby significantly improving the accuracy of anomaly detection. Thus, uncertainty and biases in water quality reporting, interpretation, and modelling are reduced, and the effectiveness of subsequent management actions improved.
\end{abstract}

%
%

\section{Introduction}
\label{sec:sample1}

Water-quality monitoring data are used to better understand and manage natural and anthropogenic impacts on water resources and aquatic ecosystems \cite{altenburger2019future}. 
In stream and river networks, the monitoring of high-frequency water-quality data often relies on digital sensors, which can be expensive, costing thousands of dollars for a single site \cite{SBIR}.
Consequently, monitoring efforts are typically localized, with sensors predominantly deployed at the system's end or outlet \cite{leigh2019predicting}.
This provides little information about water-quality dynamics in the upper reaches of the catchment, making it difficult for land managers to identify pollutant sources and/or encourage practice change from private landholders. 
However, recent advances in sensor technology have led to lower-cost options, including new sensors currently under development \cite{GreatBarrierReef}. 
These cost-effective sensors offer a valuable opportunity to collect dense spatial and temporal data from stream networks in near-real time \cite<e.g.>{isaak2017norwest}.

While there are evident advantages to employing in-situ sensors for water quality monitoring, it is crucial to acknowledge the significant challenges associated with their use . In-situ sensors are susceptible to various technical issues that can be challenging to identify and mitigate \cite{hill2010anomaly}. These issues encompass calibration errors, biofouling, and battery failure, all of which can introduce inaccuracies into the collected data. 
We refer to these as technical anomalies, but they have also been referred to as ``outliers'', ``inaccuracies'', or ``special causes of variation''. 
Some of the most common technical anomalies in water quality data include large and sudden positive or negative spikes, which are referred to as point anomalies \cite{chandola2009anomaly}. 
 Additionally, anomalies can manifest as periods of unusually low or high variability, drift, or shifts in the data, which are known as collective or persistent anomalies \cite{leigh2019framework}.
If these anomalies are not properly addressed, they can have detrimental effects on statistical inference, leading to biased estimates and impacting the spatial and temporal autocorrelation structures 
\cite{congdon2019bayesian, barnett1984outliers}. 
Therefore, it is imperative to employ robust statistical methods that can effectively distinguish technical anomalies from  natural water quality events. By doing so, high-frequency spatio-temporally dense data from in-situ sensors can be utilized to provide valuable insights into water quality dynamics for both management and research purposes.

Another challenge is that catchment water quality is spatially and temporally dynamic at intermediate and broad scales \cite{peterson2013modelling}. For example, sudden increases in water level, sediment and nutrient concentrations may be observed at multiple sites after extreme weather events (e.g. flooding, heavy rain and increased runoff), or at neighbouring time lags as water moves downstream. 
 Not surprisingly, water quality data collected in streams often exhibit spatial \cite{peterson2006patterns, garreta2010spatial, cressie2006spatial} and temporal correlation \cite{leigh2019predicting}, which tends to increase with the spatial and temporal density of data \cite{isaak2017norwest, steel2016spatial, money2009using, jackson2018spatio, santos2022bayesian}.

Generally, spatial, geo-referenced and time-series observations are considered anomalous when they differ substantially from their non-anomalous neighbours \cite{fotheringham2008sage, chandola2009anomaly}. 
The challenge in stream systems is that data exhibit unique patterns of spatial dependency due to the branching structure of the network, longitudinal connectivity, directional flow and water volume, which must be accounted for \cite{cressie2006spatial, peterson2013modelling, o2014flexible, jackson2018spatio}.

Anomaly detection in time-series, spatial, and spatio-temporal data has been extensively studied in the literature. Numerous examples can be found in various fields such as climate data analyses \cite{costa2009homogenization, resch2023quantile}, transportation \cite{shi2018detecting, djenouri2019adapted}, and intrusion detection \cite{ZHANG2020101681, wang2017hast}.
Methods such as local outlier factor (LOF) \cite{bosman2017spatial}, leave-one-out cross-validation kriging methods, extreme value theory approaches \cite{kandanaarachchi2021leave}, 
change-point detection \cite{tveten2022scalable},
and Hidden Markov models (HMM) \cite{li2017multivariate} have been proposed to identify spatial and temporal outliers. Some approaches have also been proposed to identify anomalies in datasets from stream networks where temporal dependence exist. Popular methods include autoregressive models (e.g. autoregressive integrated moving average (ARIMA) \cite{leigh2019framework}, vector autoregression models (VAR), machine learning methods (e.g. artificial neural networks (ANN), random forests,  Long Short Term Memory) \cite{rodriguez2020detecting, JONES2022105364}.)
However, little work has focused on detecting anomalous data accounting simultaneously for spatial and temporal variation in stream-network contexts, with consideration of the unique spatial relationships found in river networks.

To tackle the complex challenges in water quality monitoring, it is crucial to adopt suitable modeling frameworks that can address the following key aspects: (1) Distinguishing water quality events from technical anomalies, (2) Incorporating spatial and temporal autocorrelation to capture the underlying processes and patterns of interest more accurately, (3) Recursive update models as new data becomes available over time. 

In this paper, we present a novel spatio-temporal anomaly detection approach for water quality data obtained from arrays of sensors in stream networks. 
We introduce a pre-processing framework involving the implementation of data imputation to efficiently handle missing data, optimal alignment of time series to address disparities in time series, and the identification of water quality events to facilitate the accurate detection of anomalies in spatio-temporal data. The Bayesian modelling approach provides posterior predictive distributions for each observation, which allows the agreement between models and data to be assessed using empirical residuals.  
Statistical tools including mixture models and HMMs are then used to detect anomalies in these residuals. 
The Bayesian approach is implemented in near real-time through recursive modeling, which allows the model to update automatically as new data becomes available.
We use a comprehensive simulation study to assess and evaluate the ability of the methods to detect different types of anomalies. This is followed by a case-study involving a sensor network located in the Lower Herbert River in Far North Queensland, Australia, which discharges into the Great Barrier Reef Lagoon World Heritage Area. 

Herein we focus on technical anomalies in sensor data while taking into account water quality events. 
The approach introduced in this manuscript has been carefully developed, taking into account the specific characteristics of the case study. The proposed framework aims to effectively detect anomalies and classify {\it water quality events}, but it also has the potential to be applied to a broad range of environmental monitoring applications. By leveraging the power of advanced statistical and machine learning techniques, this methodology offers a flexible and scalable solution for monitoring various types of water quality data, which should ultimately facilitate decision-making related to the release of trustworthy data to the public domain and in the maintenance of water in-situ sensors.

\section{Motivating dataset: water quality in Far North Queensland, Australia} 

\subsection{Study region and sites}

The Herbert River basin in far north Queensland, Australia (Fig~\ref{fig:netw}) has distinct upper and lower catchments. The upper catchment, in the Southern Atherton Tableland, has an elevation over 1000m and an average annual rainfall just under 1200mm; whereas the lower catchment is a large coastal delta receiving more than double the rainfall. 
The land use in the upper part of the catchment is dominated by grazing while the lower part of the catchment is dominated by sugarcane and conservation \cite{QGMRPL}. The monitoring sites for this study were located in the Stone River sub-catchment of the Herbert River basin. This particular area is dominated by sugarcane and conservation as well as a small amount of forestry land use. 

Water quality in this area typically changes with rainfall induced by runoff from the catchments. More specifically at the start of the wet season (October to December), monitoring will observe a \emph{first flush} phenomenon where high concentrations of sediment or nutrients will occur with small waterway discharges. Later in the wet season, large events will also contribute to high concentrations of sediment \cite{QGMRPL}. However, confounding occurs if many subsequent rainfall events happen, causing the catchment to become exhausted of sediment delivery potential and increasing discharge will not necessarily result in increasing concentrations.

\subsection{Sensor installation and data collection}

\begin{figure}[htbp]
  \centering
   \includegraphics[width=4.5in]{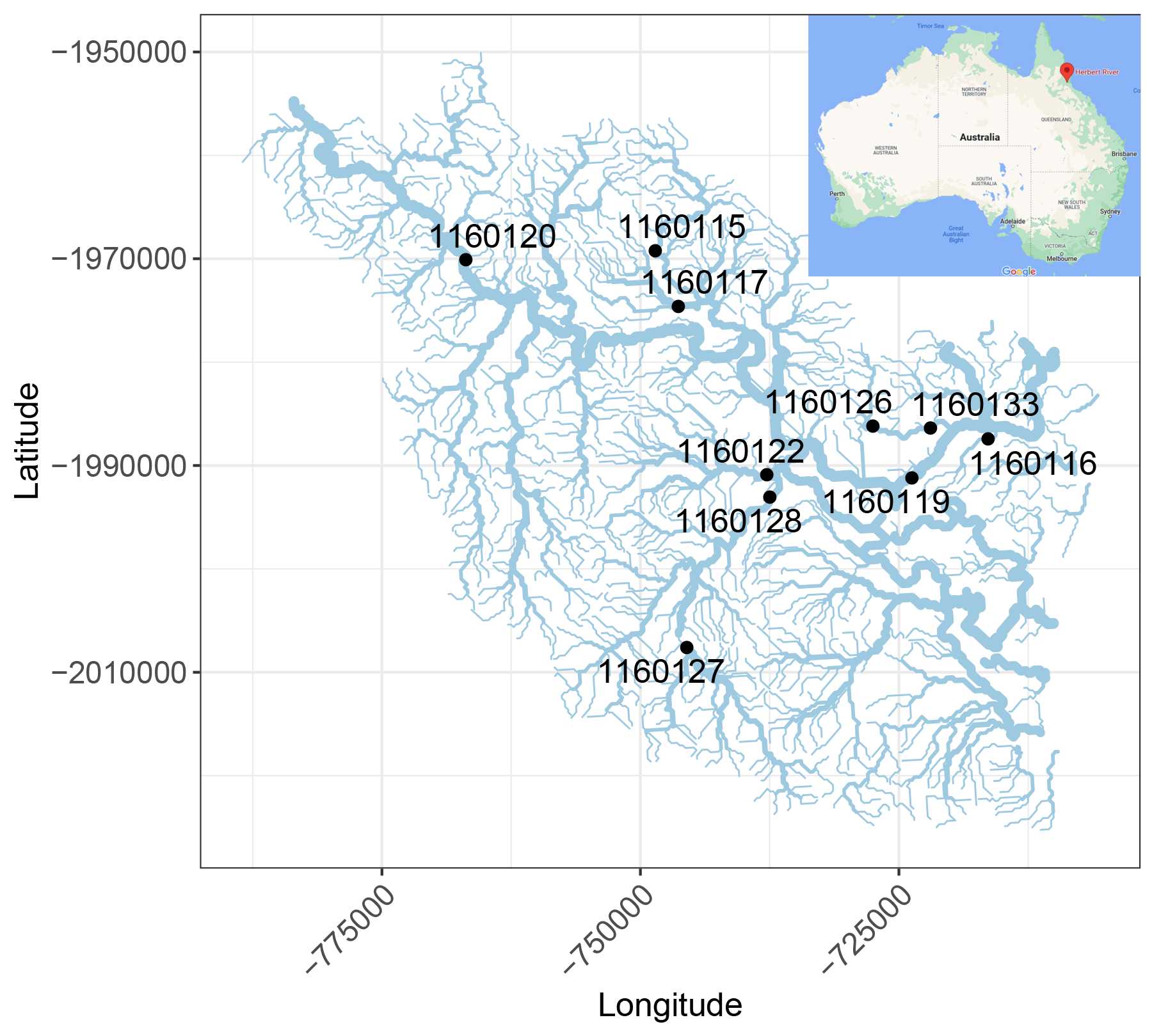}
  \caption{A sub-catchment of the Lower Herbert River in Far North Queensland, Australia. The locations correspond to 10 micro site sensors. 
   The width of each plotted stream segment in the figure is scaled in proportion to the Strahler stream order. We used Albers Equal Area projection centered at 153 degrees longitude, with standard parallels at -20 and -40 degrees latitude. }
     \label{fig:netw}
\end{figure}

\begin{figure}[htbp]
  \centering
   \includegraphics[width=5in]{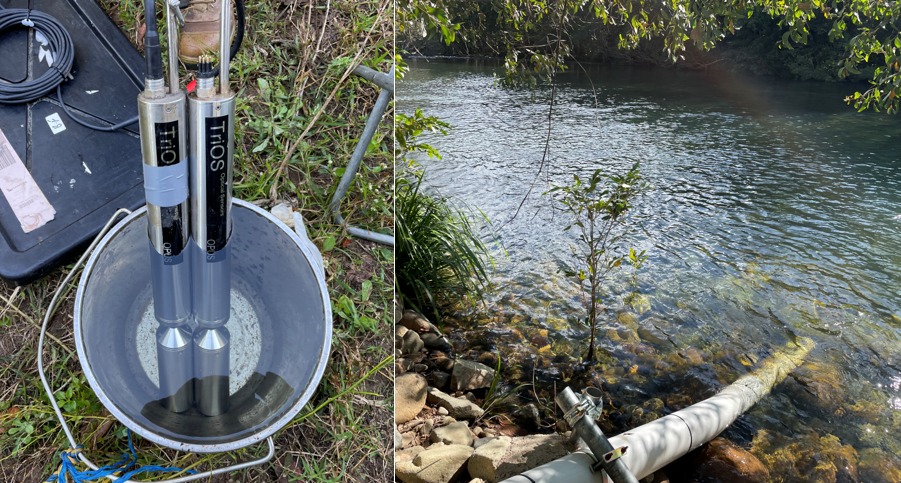}
  \caption{Opus TriOS optical water-quality sensor (model TLS02 021) produced by TriOS Mess- und Datentechnik GmbH  Opus (left) and the installation on Broadwater Creek at Day Use Area/Abergowrie State Forest in the Herbert River (site 1160115). }
    \label{fig:opus}
\end{figure}

We installed 10 Opus TriOS water-quality sensors (
Fig~\ref{fig:opus}) produced by TriOS Mess- und Datentechnik GmbH in the branching network of the Herbert River (Fig~\ref{fig:netw}). 
These Internet of Things (IoT) devices measure among other parameters stream water level (m) and spectral total suspended solids (TSSeq) (referred as TSS from now on) (mg/L) . Measurements were recorded every 30 minutes and uploaded via the 4G-CATM1 mobile network to a database. 
Each sensor was installed at the edge of the riverbank in a location best representing that part of the waterway. 
This positioning while given potential for edge effects in the data allows for service access during the majority of flow conditions while maintaining a safe distance from the bed to avoid issues arising from sedimentation. It is important to note that due to this fixed positioning, the probes are only able to measure when the water level is above the lens path of the probe.

\subsection{Anomalies in the data}

\begin{figure}[hp]
  \centering
   \includegraphics[width=6.5in]{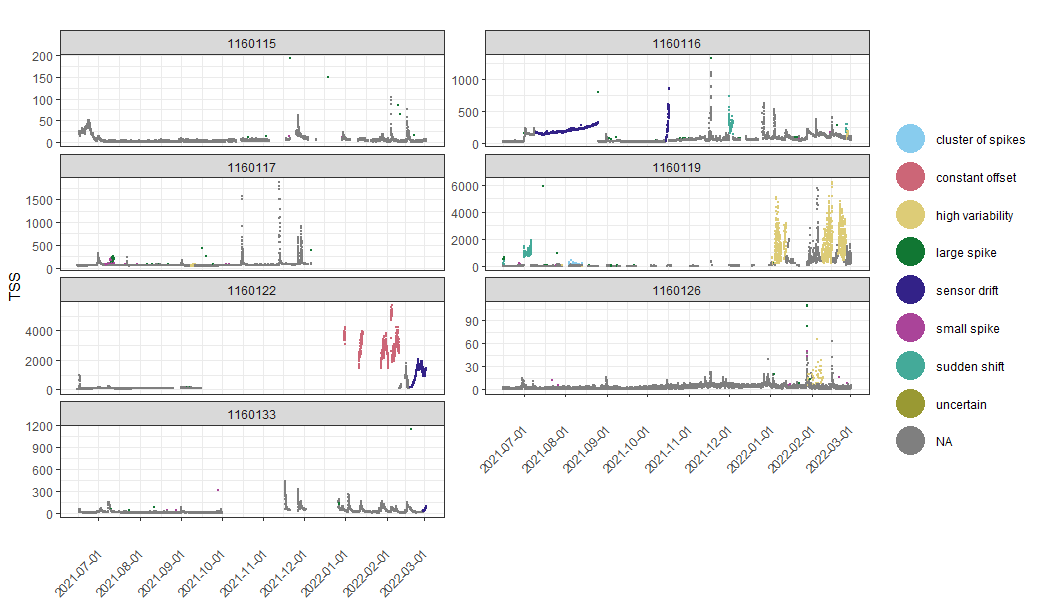}
  \caption{
  Time series of total suspended solids (TSS) data and anomalies. 
 Anomalous data points are colored according to eight types. NA refers to non-anomalous data.}
   \label{fig:turblabels}
\end{figure}

In this study, we used time-series data for TSS and water level collected between 2021-06-15 and 2022-03-02. The dataset illustrate many of the intrinsic challenges of in-situ water quality data described in the Introduction, including missing values, high variability between sites, persistent anomalies and other factors that can complicate the analysis (Figure~\ref{fig:turblabels}).
A team of water quality specialists from the Queensland Department of Environment and Science visually assessed the data and created a labelled dataset. This dataset contains labels of the technical anomalies in TSS following the framework developed by \citeA{leigh2019framework}. 
These included comparatively large and small sudden spikes (0.131\% and 0.164\% of the data, respectively), clusters of spikes (0.040\%), periods of anomalously low variability and/or persistent values and constant offsets (e.g. due to calibration errors (1.089\%)), sudden shifts (0.562\%), drifts (4.262\%) and periods of anomalously high variability (1.668\%).
The dataset also includes labels of the water events associated with rainfall (ambient conditions represent 72.308\%, while events and sensors out of the water are 14.852 and 12.841\% respectively).

The identified sensor anomalies in the dataset are generally caused by physical interference that affects the sensor readings, introducing noise and biases, thereby compromising the accuracy and reliability of the data.
Such interference can manifest as spikes in the data, which may be caused by various factors, including obstructions in the probe's lens path, such as rocks, organic debris, or living organisms like fish and other fauna.

The sensor installations leverage an attached wiper unit to clean the sensor lens from bio-fouling and other debris. However, the failure or malfunction of these units can lead to the wiper blade obstructing the lens path, causing sudden shifts or high variability in the data.  
If the wiper unit fails or the wiper blade becomes damaged, bio-fouling can accumulate on the lens, resulting in sensor drift. 
This drift refers to the gradual deviation of sensor readings from their actual values over time, which typically tends to increase in a monotonic manner.

While digital sensors are generally less susceptible to electrical interference, an insufficient power supply can result in constant values being reported due to the probe relying on memory. 
On the other hand, the level sensors used in these installations are more analog and can be susceptible to electrical interference, often due to poor grounding connections. In such instances, high variability or negative values may be produced, depending on the specific error mechanism.
Some of these anomalies are very challenging for identification using traditional methods. For instance, in location 1160116, there were two long periods of sensor drift (in purple). Similarly, in location 1160119, there was a sudden shift on the mean of the process (light green) and a period of high variability.

\section{Spatio-temporal models}

In this section, we introduce a spatio-temporal model that we use to describe stream network data.
Consider $S$ unique spatial locations on a stream network where sensors are installed. Repeated water quality measurements (e.g.\ turbidity, nitrate, conductivity or water level) are collected at all spatial locations and at regular time points $t = 1, 2 ,\ldots , T$, producing spatio-temporal stream network data. 
Let $\bm{y}_t$ be an $S \times 1$ vector of random variables observed  at time $t$ at spatial locations $s = 1, 2, \ldots S$.

 Several statistical models have been proposed to capture spatio-temporal dependence in stream network data, including those suggested by \citeA{money2009modern}, \citeA{jackson2018spatio}, \citeA{tang2020space}, and \citeA{santos2022bayesian}.
In this work, we follow the approach of \citeA{santos2022bayesian, santos2022ssnbayes} where the following vector auto-regressive spatial linear model is used to describe potential spatio-temporal dependencies in steam data.

\begin{equation}
[\bm{y}_t \mid  \bm{y}_{t-1},\bm{X}_{t},\bm{X}_{t-1},\bm{\beta}, \bm{\Phi}_1, \bm{\Sigma}, \sigma^2_0 ] = \mathcal{N}(\bm{\mu}_{t},\bm{\Sigma} + \sigma^2_0\bm{I}),  
\label{eq:eq1}
\end{equation}

\begin{equation}
\bm{\mu}_{t} = \bm{X}_{t}\bm{\beta} + \bm{\Phi}_1 (\bm{y}_{t-1} - \bm{X}_{t-1}\bm{\beta}),
\label{eq:err_}
\end{equation}

\noindent where $\bm{y}_{t-1}$ is the process at time $t-1$,
$\bm{X} = [\bm{X}_1, \bm{X}_2,\cdots,\bm{X}_T]$ where each $\bm{X}_t$ is a $S \times p$ design matrix of $p$ covariates at time $t$, and
$\bm{\beta}$ is a vector of regression coefficients of length $p$. 
The total spatial variance-covariance of $\bm{y}_t$ can be written as $\bm{\Sigma} + \sigma^2_0\bm{I}$.
Here $\sigma^2_0\bm{I}$ incorporates the unstructured error term, where $\sigma_{0}^2$ is known as the nugget effect and $\bm{I}$ is the identity matrix.  The notation $\mathcal{N}(·,·)$ is the probability density function for the multivariate normal distribution.

In Equation (\ref{eq:err_}), the mean $\bm{\mu}_t$ is expressed as a vector autoregression of order one, VAR(1), where $\bm{\Phi}_1$ is a $S \times S$ square transition matrix with diagonal parameters $\phi_s$ that determine the magnitude of first-order temporal autocorrelation. Note that this model can be extended to include more complex autocorrelation terms, for further details see \citeA{santos2022bayesian}.
By setting $\bm{\Sigma}$ and $\bm{\Phi}_1$ to zero and neglecting the temporal and spatial correlation, we can derive the conventional linear regression model.

Studies in the literature have investigated spatial correlation in stream data, with research conducted among others by  \citeA{cressie2006spatial, ver2006spatial, money2009using, garreta2010spatial, steel2016spatial, jackson2018spatio, tang2020space}.
These approaches allow information to be borrowed from neighboring sites, which is crucial in identifying anomalies in spatio-temporal data.
We can express the spatial covariance matrix $\bm{\Sigma}$ of dimension $S \times S$ using a mixture approach \cite{ver2010moving}, that describes multiple patterns of spatial autocorrelation often found in data collected on streams:

\begin{equation}
\bm{\Sigma} = \bm{C}_{ED} + \bm{C}_{TU} + \bm{C}_{TD}   
= \sigma_e^2\bm{R}_{e}(\alpha_e)+\sigma_u^2\bm{R}_{u}(\alpha_u) + \sigma_d^2\bm{R}_{d}(\alpha_d). 
\label{eq:covs}
\end{equation}

\noindent Here, $\bm{C}_{ED}$ is a standard geostatistical covariance matrix based on Euclidean distance. The terms $\bm{C}_{TU}$ and $\bm{C}_{TD}$ are tail-up and tail-down covariance matrices based on stream distance.

The parameters $\sigma_e^2$,  $\sigma_u^2$ and  $\sigma_d^2$ are the partial sills, and analogous parameters $\alpha_e$, $\alpha_u$, $\alpha_d$ are the spatial range parameters. Similarly,  $\bm{R}_{e}$, $\bm{R}_{u}$ and $\bm{R}_{d}$ are the correlation matrices. Several variations of this model can be fit using the R statistical software \cite{rprog} package \textsf{SSNbayes} \cite{santos2022ssnbayes}.
Here we fit a fully Bayesian model in Stan \cite{rstan} to estimate the posterior distribution via MCMC. 
A reproducible example demonstrating the use of the 
\textsf{SSNbayes} package for anomaly detection is available on \url{https://www.kaggle.com/code/edsans/anomaly-detection-using-ssnbayes/notebook}.

Tail-up/tail-down covariance models are formulated by integrating a kernel function over a white noise process that is strictly upstream/downstream of a given site. The tail of the moving-average function points upstream in the tail-up model, and downstream in the tail-down model. 
In the tail-up model, autocorrelation is confined to sites that are connected by flow, while in the tail-down model, spatial dependence can still exist even between sites that are not directly connected by flow.
In the tail-up and tail-down models, the distance is restricted to the stream network, which captures unique spatial relationships produced by the branching network structure, longitudinal connectivity, flow volume and flow direction of stream networks. See \cite{ver2006spatial, ver2010moving} for more details. 

The estimation of spatial covariance matrices involves estimating the partial sill and range distributions. To model these parameters, we employ flat uniform priors for both spatial components. Additionally, we set an upper bound of four times the maximum distance between observations in the river network for the range parameter. The temporal structure is defined by the matrix $\bm{\Phi}$, and we adopt a uniform prior ranging from -1 to 1 to model the diagonal autoregressive parameter.

\subsection{Bayesian approaches for spatio-temporal data}

To implement the above model within a Bayesian framework we use the Bayes theorem, which involves the computation of the posterior distribution for the parameters of interest $\bm{\theta} = \{\bm{\beta}, \alpha, \sigma_{.}, \bm{\Phi}, \sigma_0\}$:

\begin{equation}
f\left (  \bm{\theta} |{\bf y} \right ) = \frac{f\left ( {\bf y} | \bm{\theta} \right ) f\left ( \bm{\theta}  \right ) }{
\int f\left ( {\bf y} | \bm{\theta}  \right ) f\left (  \bm{\theta}  \right ) d \bm{\theta}},
\label {eq:eqf3}
\end{equation}

\noindent  where $f\left ( \bm{\theta}  \right )$ represents the prior distribution for $\bm{\theta} $ and $f\left ( {\bf y} | \bm{\theta}  \right )$ is the likelihood of the data $\bm{y}$ given $\bm{\theta} $.

Model goodness-of-fit can be assessed by comparing the data ($\bm{y}_t$) to the predictions obtained via posterior predictive distributions, which measure the agreement between the assumed model and the observed data. The ability to identify observations that significantly deviate from the predictions is especially useful for detecting anomalies in the data \cite{gelman2006data}. The fitted vector autoregression spatial stream-network model is used to generate the posterior predictive distribution (PPD), ($\hat{\bm{y}_t}$), which is obtained from the observed response variable ($\bm{y}_t$) and covariates for the observed and predicted data ($\bm{X}_t$ and $\hat{\bm{X}_t}$) by marginalising over the likelihood as follows:

\begin{equation}
p(\hat{\bm{y}_t} \mid \bm{y}_t, \bm{X}_t, \hat{\bm{X}_t}) = 
\int p(\hat{\bm{y}_t} \mid \bm{\theta}, \hat{\bm{X}_t}) p(\bm{\theta} \mid \bm{y}_t, \bm{X}_t) \, \textrm{d}\bm{\theta},
    \label{eq:pred}
\end{equation}

\noindent where $\bm{\theta}$ is a vector of estimated parameters from the fitted model (i.e. posterior distributions). 
Note that the vector $\bm{y}_t$ may contain missing values, which may occur for various reasons.  
These missing values are imputed using the posterior predictive distributions integrated into our Bayesian inference approach.

 The vectors $\bm{X}_t$ and $\hat{\bm{X}_t}$ are the same when we produce predictions for the observed data points. The model prediction error or empirical residual for an observation $y_{s,t}$ at location $s$ and time $t$ is then $e_{s,t} = y_{s,t} - \hat{y}_{s,t}$, where $\hat{y}_{s,t}$ is calculated based on a summary of location of the posterior predictive distribution such as the posterior mean.

\section{Anomaly detection methods}

Several statistical approaches can be used to detect anomalies in stream data once a spatio-temporal model is fit, and the posterior predictive distribution and the prediction errors have been computed using the estimated parameters. Here we consider four different approaches including the 1) posterior predictive approach, 2) finite mixture approach, 3) hidden Markov modelling approach, and 4) autoregressive integrated moving average method. Importantly, these approaches are all considered unsupervised, meaning that they do not require a dataset containing labelled anomalies and non-anomalous/normal data.
Unsupervised models provide multiple benefits, since labelled datasets are often unavailable for new data sources or monitoring sites, and are time consuming to generate manually. In addition, labelled datasets may not be transferable to new sites or time periods if they have been generated under different environmental conditions. 
\\

\noindent { \bf   Method 1: Posterior predictive approach} 

The first approach we introduce is obtained via the posterior predictive distribution from Equation (\ref{eq:pred}). 
 Specifically, this approach is reasonable as it compares the predicted values from the model with the associated uncertainty to the observed data, and is based on the assumption that anomalous observations will have a low probability under the fitted model \cite{murphy2012machine}.  
This method is expected to be effective in identifying spikes in the data and sudden changes in the mean of the process.
The steps to implement this approach are as follows:
\\

\noindent Steps: 

\begin{enumerate}
    \item We define anomalous observations as those that fall outside the 95\% highest posterior density or posterior prediction interval (HDI). This significance level will be utilized in both the simulation and the case study. 
    Note that tighter limits should be used if the data are expected to have a large proportion of anomalies.
    \item Exclude these anomalous observations from the data. 
    \item 
    In one of our approaches we iterate steps 1-2 twice. Anomalies identified in both the first and second iterations are combined, flagged, and used to compute the performance measures.
\end{enumerate}

For this study, we considered up to two iterations, to ensure that we did not exclude too many observations, but other alternatives are possible \cite{weisberg2013applied}.
Our first approach (iter\_1) only involves fitting the model and no data are excluded (i.e. only uses step 1); whereas in the second case (iter\_2), the model is fitted again, after removing the anomalous observations (steps 1, 2, 3 and 1). 
Note that this is in accordance with control chart methodologies where existing outliers are removed while estimating the control limits  \cite{montgomery2020introduction}. 
This method considers the uncertainty in the process and can produce a probabilistic score for each observation being an anomaly, which allows observations to be flagged (e.g. data not included in the 80\% HDI).
\\

Next, we concentrate on a range of statistical methods based on the residual $e_{s,t}$ obtained using the mean of the posterior distributions. However, these approaches can be extended to use the samples from the Markov chain Monte Carlo (MCMC) simulations. 
 That is, the proposed approach would enable the implementation of fully Bayesian anomaly detection methods, utilizing all the samples drawn from the posterior distribution instead of relying on the point estimates of the residuals. \\

\noindent { \bf   Method 2: Finite mixture approach} 

Mixtures are statistical models that are particularly useful to describe response variables from a population composed of different groups or is generated under different conditions \cite{gelman1995bayesian}. 
Mixtures and other clustering-based methods have been widely used to detect anomalies  \cite{chandola2009anomaly}. 
In these settings, we assess heterogeneity in the residuals to identify anomalous data, especially when shifts in the mean of the process occurs. 
Latent observed groups in the residuals could indicate the presence of anomalous data.
Anomalous data are expected to fall outside the range of the typical distribution and reside in the tails, indicating the existence of different latent classes. 
These anomalous groups may have different statistical properties compared to the non-anomalous data, such as a higher variance or a different statistical distribution. \\

\noindent To implement this approach, we propose the following steps: 

\begin{enumerate}
    \item  First, we need to determine the appropriate number of clusters or components (e.g. $K = 3$). This can be done by inspecting the distribution of the residuals. For example, three components could be used if we expect anomalies above and below the mean. 
    \item Let $z_{s,t}$ be the cluster membership of the observed residual $e_{s,t}$ at location $s$ and time $t$, and $z_{s,t} \sim \textrm{categorical}(1/K)$ and $Z$ denotes the number of clusters. 
    \item Fit the Bayesian mixture model $e_{s,t} | z_{s,t} \sim \mathcal{N}(\mu_{z_{s,t}},  \sigma_{z_{s,t}})$, which produces a cluster membership as well as estimates of the mean and variance for each cluster.
    \item   Inspect the resulting clusters to identify anomalous components, which are typically characterized by observations found in the tails of the residual distribution and are likely to contain fewer observations than the main cluster(s) representing the non-anomalous data.
    
\end{enumerate} 

\noindent { \bf   Method 3: Hidden Markov model approach} 

In a hidden Markov model (HMM), the response variable is determined by a hidden Markov process, where the state at time $t$ depends only on the state at the previous time point $t-1$ \cite{zucchini2009hidden}. 
The HMM model can be defined by $\delta = \left(A, B, \pi\right)$, where $A$ and $B$ are the transition probability matrix and emission probability matrix, respectively, and $\pi$ is the initial state probability. 
The transition probability matrix
gives the probability distribution of observed variables given the hidden states of the model, while the emission probability matrix describes the probability distribution of observed variables conditional on the state.

This model is often referred to as dependent or dynamical mixtures and is employed with time series data.
Our proposed modeling framework involves the consideration of two or more unknown latent states ($L$), such as normal and anomalous, and we seek to compute the probability of each observation belonging to these states. 
Under this model, it is assumed that the process transitions between the latent states using a transition probability matrix. 

Conditional on the state $l_t$ , the distribution of the residual $e_t$ is assumed to be normally distributed with a mean $\mu_{l_t}$ and a variance $\sigma_{l}^{2}$.
The HMM approach is commonly used in situations where data come from multiple latent states, such as non-anomalous and anomalous states.

\begin{equation}
    e_t | l_t= \mathcal{N}(\mu_{l_t},  \sigma_{l}^{2}),
\label{eq:hmm} 
\end{equation}
 
\noindent where  $\mathcal{N}(·,·)$ is the probability density function of the normal distribution. \\

\noindent To implement this approach, the following two steps are used:  

\begin{enumerate}
    \item Fit a HMM model with $L = 2$ or more components.
    \item Use the component's membership to identify potential anomalies. 
\end{enumerate}

\noindent {
Our method is implemented using a Bayesian approach, which assigns each observation a probability of belonging to either the anomalous or non-anomalous cluster. This approach provides flexibility in the allocation of observations to different states. We have also incorporated the assumption of unequal variances between the clusters.
} \\

\noindent { \bf   Method 4: Autoregressive integrated moving average (ARIMA) (Benchmark method):}

An ARIMA approach \cite{leigh2019framework} is purely temporal, meaning that it does not account for spatial correlation, and is used to benchmark the spatio-temporal methods proposed here. 
An appropriate ARIMA model was obtained for each time series using the \textsf{auto.arima} function in R from the \textsf{forecast} package \cite{forecast} which finds the best model based on AIC. Observations were considered anomalous if they fell outside of the 95\% prediction interval. 
Following \citeA{leigh2019framework}, the general ARMA model can be described as:

\begin{equation}
\bm{y}_t  = c + \phi_1 \bm{y}_{t-1} + \cdots + \phi_m \bm{y}_{t-m} + \theta_1 \epsilon_{t-1} + \cdots + \theta_q \bm{\epsilon}_{t-q} + \bm{\epsilon}_t
\end{equation} 
\noindent where $c$ is a constant term, 
$\phi_1, \dots, \phi_p$ are scalar autoregressive coefficients of the $m$ terms, $\theta_1, \dots, \theta_q$ are the coefficients of the $q$ moving average terms $\epsilon_{t-1}, \dots, \epsilon_{t-q}$ are the errors at times $t-1$ through $t-q$.

\section{Simulation study}

The methods previously presented are largely unexplored in the detection of technical anomalies in spatio-temporal stream network data. Hence, we performed a simulation study to assess the suitability of methods 1-4 under several scenarios found in real-world applications including in our case study.
\\

\begin{figure}[ht]
  \centering
   \includegraphics[width=6in]{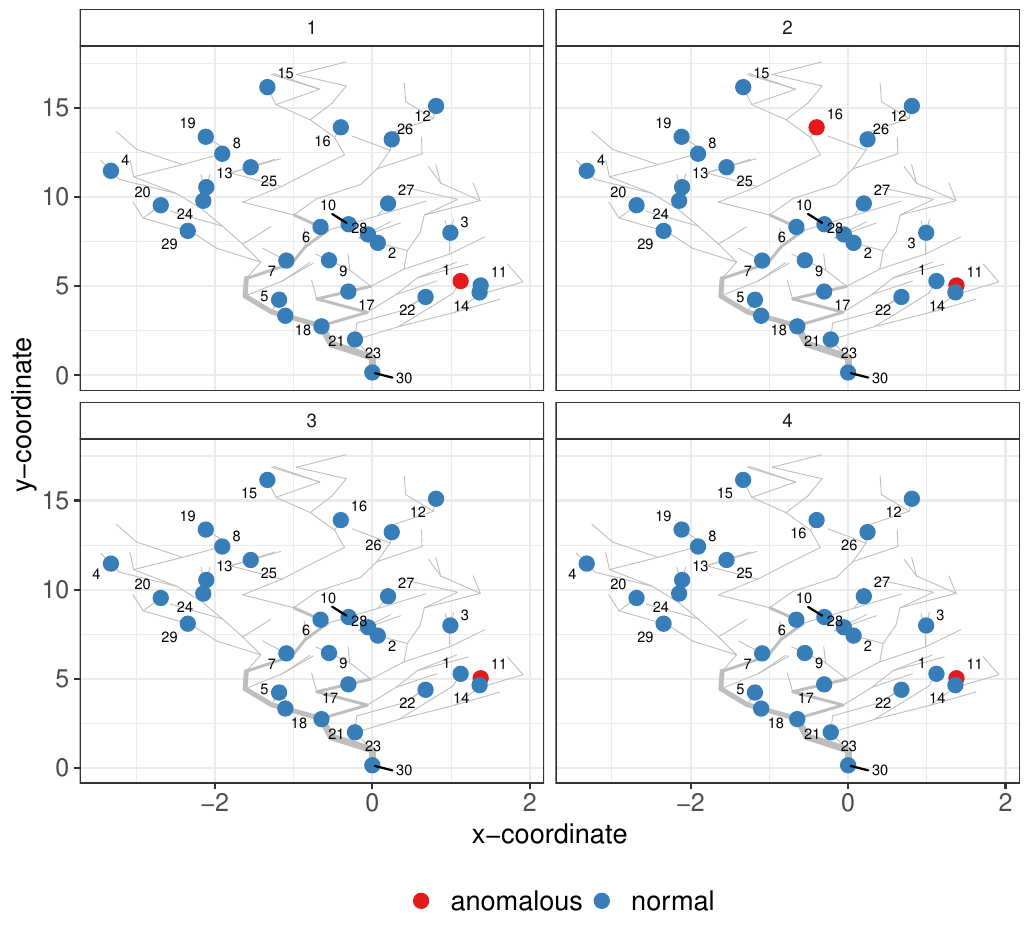}
  \caption{The artificial stream network with 30 spatial locations and observations at the first four time points. Technical water quality anomalies are shown in red and normal data are blue. The relative flow volume for each stream segment is proportional to the width of the gray line.}
  \label{fig:network}
\end{figure}

To begin, we generated simulated datasets using the following steps:

\begin{enumerate}
\item 
The installation and maintenance of TSS and nitrate sensors entail a considerable financial cost, and consequently, monitoring programs in Queensland, Australia, typically have fewer than 30 sensors per catchment. Despite this limited number of sensors, it is still possible to achieve a reasonable and realistic spatial coverage that exhibits spatial correlation in the network. 
The use of a limited number of sensors is a common scenario in environmental monitoring, and it poses several challenges such as optimal sensor placement and data sparsity issues.
Accordingly, we created artificial stream networks made up of 150 segments and $S = 30$ spatial locations representing water quality sensors. This was undertaken via using the function \textsf{createSSN()} in the \textsf{SSN} R software package \cite{hoef2014ssn}.
Fig~\ref{fig:network} shows a simulated stream network with 30 spatial locations across the first four time points.
\item We selected a common covariance structure to model spatial dependence by means of a exponential tail-down model. 
To specify this structure, we used the following spatial parameters: partial sill, which represents the variance after accounting for the nugget effect, with a value of $\sigma^2_{TD}=3$ (this value was selected based on case study from  \cite{santos2022bayesian}); the spatial range (defines the rate of decay in the covariance as a function of the separation distance) $\alpha_{TD}=10$ (see Equation (\ref{eq:covs})), and nugget effect $\sigma^2_{0} = 0.1$ (Equation (\ref{eq:eq1})).
\item We considered three covariates for the mean response which were arbitrarily generated from three standard normal covariates plus an intercept ($p = 4$) at every spatial location with regression coefficients:  $\{\beta_1, \beta_2, \beta_3\} =\{ 1, 0, -1\}$ for the slopes and the intercept $\beta_0 = 10$.
\item The spatio-temporal data were generated additively: 
     \begin{enumerate}
         \item[ a)] First, a response variable $y$ was simulated from a spatial stream-network process (at time point $t = 1$) using $\bm{y} = \bm{X}\bm{\beta} + \bm{v}$, where 
         $\bm{v} \sim N(0, \Sigma)$.
         \item[ b)] Next, we defined a temporal process based on an autoregressive function of order one, AR(1), with $\phi = 0.80$ and included $T = 120$ time points. We generated the time series data using the \textsf{corAR1} function from the \textsf{nlme} package \cite{nlme}. 
         \item[ c)] We added the temporal variation (plus some random variation $\mathcal{N}(0,1)$ ) to the spatial process simulated in (a), resulting in a spatio-temporal stream network dataset. 
     \end{enumerate}
\item To create anomalies, we first defined $q_{ini} = 0.05$, which represents the {\it initial} probability of an observation at time $t$ at a spatial location $s$ being an anomaly. 
Within the time series, a binary indicator $d$ representing the start of an anomaly was created at each location $s$ and time $t$: 

\begin{equation}
d_{st} \sim \textrm{Bern}(q_{ini}) \begin{cases}
0 & \textrm{= non-anomalous,} \\ 
1 & \textrm{= anomalous} .
\end{cases}
\label{eq:dd}
\end{equation}

\noindent The binary indicator is also shown in Fig~\ref{fig:network} in the spatial context of the network.

\item We considered four types of anomalies according to observed data in the case study introduced in the motivating dataset and reported in the literature \cite{leigh2019framework}.
This included (1) moderate/large sudden spikes, (2) anomalously high variability, (3) a shift in the mean, and (4) drift; each with the same initial probability of occurrence. 
\item For each $d_{s,t} = 1$ we simulated the persistence of the anomaly (i.e. number of consecutive observations with anomalous data) using a Poisson distribution  $n_d \sim Poisson(\lambda)$ with rate $0.8$, which determines subsequent anomaly indicators $d_{s,t+1}$, $d_{s,t+2}$, etc.
Persistent anomalies occur for high variability, shift, and drift anomalies, while spikes are only composed of one observation at a single time point.
\item We defined the observed response variable, $y_{obs}$, as a mixture distribution:

\begin{equation}
y_{obs_{st}} = \left\{\begin{matrix}
y_{st} + d_{s,t} \mathcal{N}(5,1) & \textrm{for spikes,}  \\ 
y_{st} + d_{s,t} \mathcal{N}(1,5) & \textrm{for high variability,} \\
y_{st} + d_{s,t} \mathcal{N}(5,1) & \textrm{for shift,} \\
y_{st} + d_{s,t} (\mathcal{N}(5,1)) & \textrm{for drift,}  
\end{matrix}\right.
\label{eq:dd2}
\end{equation}

\noindent  where (.) represents sort ascending to create a drifting pattern.  

\end{enumerate}

\begin{figure}[ht]
  \centering
   \includegraphics[width=6.5in]{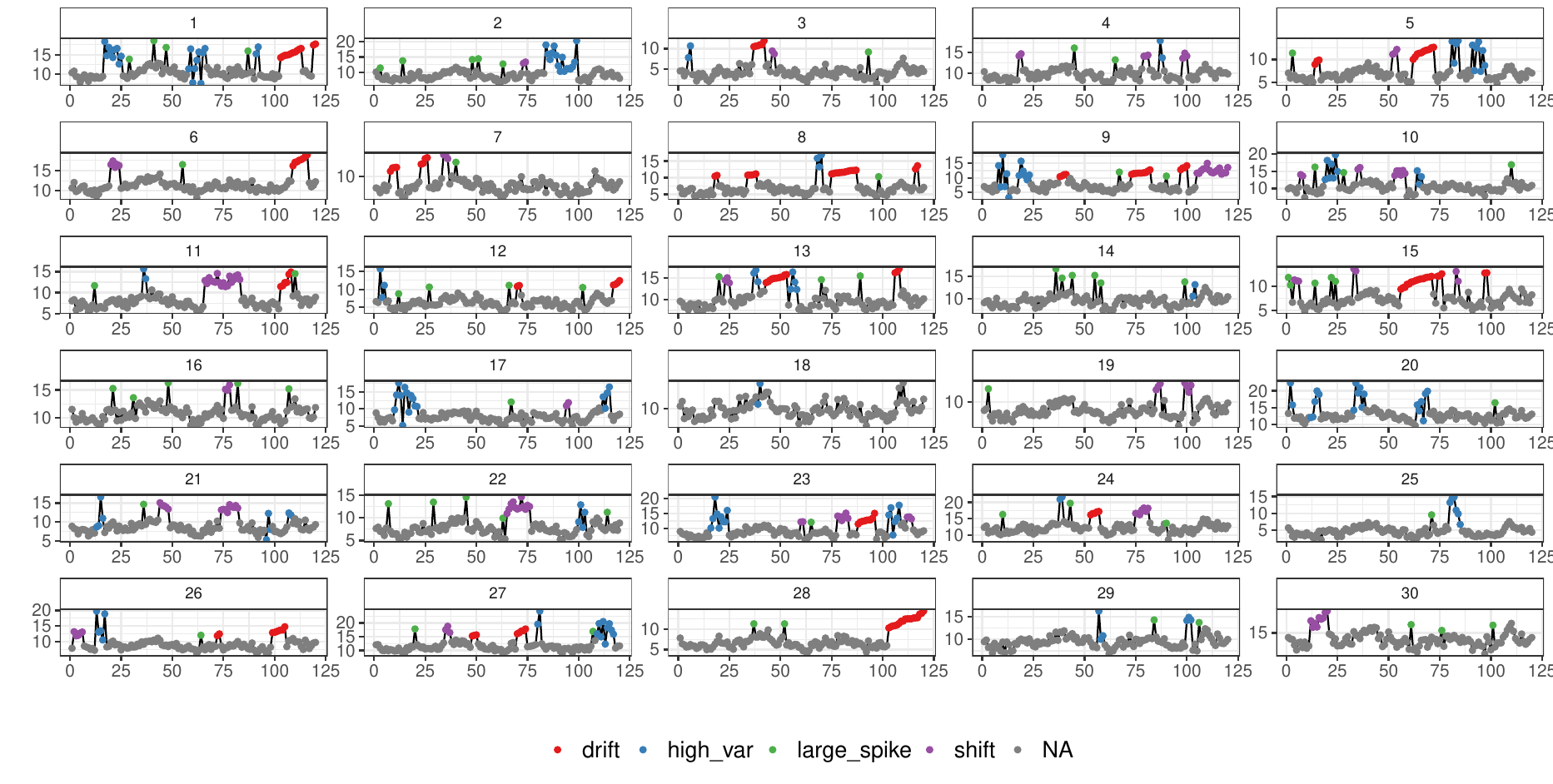}
  \caption{Time series of anomalous and normal observed data at 30 locations, across the 120 time points. Anomalous data  represent drift, high variability, large spikes, and sudden shifts and non-anomalous data are labelled as NA.} 
  \label{fig:ts}
\end{figure}

A time series representation of the simulated data is shown in Fig~\ref{fig:ts}, where non-anomalous data are represented in gray and the different anomaly types are represented with red, purple, green and blue dots. 
Notice that some anomalies were one-off cases (e.g. spikes in location 28), while others persisted through time at a single location (e.g. drift in location 28).
R codes to reproduce the simulations are part of the supplementary materials.

\subsection{Simulation results}

After fitting the spatio-temporal model from Equation (\ref{eq:eq1}), we used approaches 1-4 (PPD, mixtures,  HMMs, ARIMA) to detect the anomalies in the simulated dataset. Two iterations were considered for the PDD and the finite mixtures. 
We simulated 100 datasets and computed the performance using a variety of model measures across the different methods.

The sensitivity of the method ($se = TP/(TP + FN)$) measures the ability to detect the true positive ($TP$) anomalies when they are present. Here, false negatives ($FN$) are truly anomalous observations classified as non-anomalous. 
We also calculated the specificity ($sp = TN/(TN + FP)$), which measures the ability of the method to correctly identify normal (non-anomalous) data when anomalies are absent ($TN$). The false positives ($FP$) occur when non-anomalous observations are classified as anomalous. 
The overall accuracy of the method ($acc = (TP + TN)/(TP + TN + FN+ FP)$)  measures the proportion of correctly identified observations. However, the data are unbalanced (i.e. few anomalies compared to normal data) and so we also employed the adjusted or balanced accuracy ($acc_{adj}$) \cite{brodersen2010balanced} as the midpoint between the $se$ and the $sp$. Finally, we calculated the Matthews correlation coefficient ($MCC = \frac{TP \cdot TN - FP \cdot FN} {(TP + FN) \cdot (TP + FP) \cdot (TN + FP) \cdot (TN + FN)}$) \cite{matthews1975comparison}, which makes use of all the elements of the confusion matrix (containing  the number of correct and incorrect predictions made by the model) to measure the quality of a binary classification.

Model overall performance measure results are provided in Table~\ref{table:perfMeas2}.
All of the methods outperformed the benchmark (i.e. ARIMA) in terms of accuracy, adjusted accuracy, $MCC$ and $se$.
These results indicate that our spatio-temporal approaches have a higher efficacy in detecting anomalies compared to the benchmark method.
The PPD method tended to have lower sensitivity than the finite mixtures and the HMM. This could be caused due to the bias in the parameters produced by the anomalies. Removing the anomalies and refitting the model, tended to produce better performance overall in terms of MCC.
The PPD using two iterations and HMM were the best individual performing method in terms of MCC. 

Persistent anomalies, which occur at multiple time points, were challenging for all of the methods to detect because they affect the estimated parameters (i.e. temporal correlation), causing bias and distorting the predictions. However, these models captured well the beginning and the end of the anomalous period, which may be useful for manual quality coding or labelling the whole period as abnormal by water quality specialists.

We calculated performance measures by method and anomaly type to determine whether the methods had a stronger ability to detect some anomaly types more than others (See Table \ref{table:perfMeasANom}. 
Large spikes were identified by all the approaches with high sensitivity, especially by the mixtures ($se > 0.99$)
The PPD seemed to have less propensity for false positives ($1 - sp$) and the highest $MCC$.
Anomalies that involved high variability in the data were better identified using HMM based on MCC and adjusted accuracy. When shifts in the mean occurred, the HMM also tended to outperform the rest of the methods. 
Drift was by far the hardest type of anomaly to identify. Apart from the finite mixtures, the rest of the methods had a low probability of detecting these subtle patterns in the data (low $se$). Nevertheless, most of the models tended to detect the beginning of the drift period, which could be used to automatically flag these patterns in the dataset and then manually adjust them. 
In terms of model computational complexity, the PPD (iter\_1) method is relatively low in complexity since it involves fitting the spatio-temporal model only once. The PPD (iter\_2) model, on the other hand, has a medium complexity as it requires fitting the models twice after excluding anomalies in the first iteration. 
Mixtures and the HMM exhibit high complexity due to their reliance on residual computation in the spatio-temporal model and the inherent challenges associated with handling identifiability issues. 
This qualitative aspect is critical from both the implementation and operational perspectives. It guides the decision-making process toward simpler models when all methods (PPD, mixtures, and HMM) perform well.

\begin{table}[]
    \caption{A comparison of model performance measures and qualitative factors using ARIMA (benchmark), PPD (two iterations), finite mixtures (two iterations) and the HMM models. Performance measures include sensitivity (\textit{se}), specificity (\textit{sp}), accuracy (\textit{acc}), adjusted accuracy (\textit{$acc\_adj$}), and the Matthews correlation coefficient (\textit{MCC}). 
    Model computational complexity is a subjective measure based on our perception.    } 
    \label{table:perfMeas2}
    \centering
    \begin{tabular}{lllllll}
    Approach & $se$ & $sp$ & $acc$ & $acc\_adj$ & $MCC$ & Complex\\ \hline
    ARIMA & 0.3390 & 0.9485 & 0.8690 & 0.6437 & 0.3388 & {\bf low} \\ 
  PPD (iter\_1) & 0.3593 & {\bf 0.9707} & 0.8921 & 0.6650 & 0.4272 & {\bf low} \\ 
  PPD (iter\_2) & 0.4849 & 0.9616 & {\bf 0.9002} & 0.7232 & 0.5064 & med \\ 
  Finite mixtures (iter\_1) & {\bf 0.8896} & 0.6479 & 0.6787 & 0.7687 & 0.3638 & high \\ 
  Finite mixtures (iter\_2) & 0.8749 & 0.6888 & 0.7124 & {\bf 0.7818} & 0.3874 & high \\ 
  HMM & 0.5587 & 0.9450 & 0.8953 & 0.7519 & {\bf 0.5197} & high\\ \hline
    \end{tabular}
\end{table}

\begin{figure}[ht]
  \centering
   \includegraphics[width=5in]{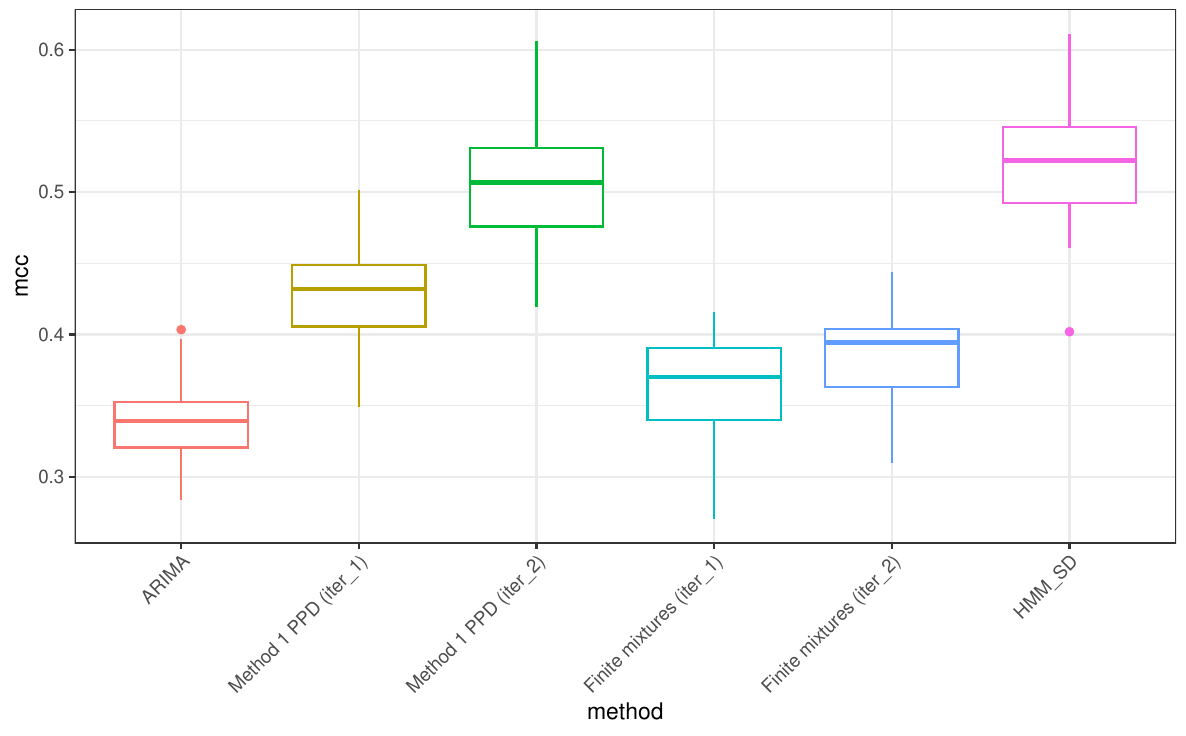}
  \caption{Comparison of the Matthews correlation coefficient (MCC) in 100 simulated datasets. }
  \label{fig:mcc}
\end{figure}

\begin{table}[]
    \caption{Performance measures by anomaly type, with models (ARIMA, PPD with two iterations, finite mixtures with two iterations and the HMM models) listed in increasing level of complexity. The best performing method in terms of MCC are highlighted in bold.}
    \label{table:perfMeasANom}
    \centering
    \begin{tabular}{lllllll}
   Anomaly & Approach & $se$ & $sp$ & $acc$ & $acc\_adj$ & $MCC$\\ \hline

  large\_spike & ARIMA & 0.8077 & 0.9489 & 0.9459 & 0.8783 & 0.4360 \\ 
  large\_spike & Method 1 PPD (iter\_1) & 0.8183 & {\bf 0.9709} & {\bf 0.9675} & 0.8946 & 0.5493 \\ 
  large\_spike & Method 1 PPD (iter\_2) & 0.9315 & 0.9616 & 0.9609 & {\bf 0.9466} & {\bf 0.5595} \\ 
  large\_spike & Finite mixtures (iter\_1) & {\bf 0.9939} & 0.6468 & 0.6544 & 0.8203 & 0.1951 \\ 
  large\_spike & Finite mixtures (iter\_2) & 0.9888 & 0.6885 & 0.6951 & 0.8386 & 0.2120 \\ 
  large\_spike & HMM & 0.9479 & 0.9454 & 0.9454 & {\bf 0.9466} & 0.5027 \\ \hline

  high\_var & ARIMA & 0.4258 & 0.9489 & 0.9276 & 0.6874 & 0.2941 \\ 
  high\_var & Method 1 PPD (iter\_1) & 0.4534 & {\bf 0.9709} & {\bf 0.9503} & 0.7122 & 0.3920 \\ 
  high\_var & Method 1 PPD (iter\_2) & 0.5590 & 0.9616 & 0.9455 & 0.7603 & 0.4271 \\ 
  high\_var & Finite mixtures (iter\_1) & 0.7004 & 0.6468 & 0.6488 & 0.6736 & 0.1401 \\ 
  high\_var & Finite mixtures (iter\_2) & {\bf 0.7826} & 0.6885 & 0.6920 & 0.7355 & 0.1943 \\ 
  high\_var & HMM & 0.6489 & 0.9454 & 0.9336 & {\bf 0.7971} & {\bf 0.4309} \\ \hline

   shift & ARIMA & 0.2081 & 0.9489 & 0.9142 & 0.5785 & 0.1383 \\ 
  shift & Method 1 PPD (iter\_1) & 0.2335 & {\bf 0.9709} & {\bf 0.9372} & 0.6022 & 0.2178 \\ 
  shift & Method 1 PPD (iter\_2) & 0.3824 & 0.9616 & 0.9349 & 0.6720 & 0.3114 \\ 
  shift & Finite mixtures (iter\_1) & {\bf 0.9173} & 0.6468 & 0.6590 & {\bf 0.7820} & 0.2404 \\ 
  shift & Finite mixtures (iter\_2) & 0.8260 & 0.6885 & 0.6946 & 0.7572 & 0.2249 \\ 
  shift & HMM & 0.4766 & 0.9454 & 0.9237 & 0.7110 & {\bf 0.3334} \\ \hline

drift & ARIMA & 0.1592 & 0.9489 & 0.9194 & 0.5541 & 0.0879 \\ 
  drift & Method 1 PPD (iter\_1) & 0.1578 & {\bf 0.9709} & {\bf 0.9413} & 0.5643 & 0.1308 \\ 
  drift & Method 1 PPD (iter\_2) & 0.2795 & 0.9616 & 0.9366 & 0.6206 & 0.2090 \\ 
  drift & Finite mixtures (iter\_1) & {\bf 0.9909} & 0.6468 & 0.6593 & 0.8189 & 0.2450 \\ 
  drift & Finite mixtures (iter\_2) & 0.9704 & 0.6885 & 0.6988 & {\bf 0.8294} & {\bf 0.2595} \\ 
  drift & HMM & 0.3450 & 0.9454 & 0.9234 & 0.6452 & 0.2185 \\ \hline

    \end{tabular}
\end{table}

The posterior distributions of the regression coefficients obtained from the fitted model indicated that all the regression parameters were slightly biased, apart from $\beta_1$ which captured the true value of 1 (Fig~\ref{fig:betas}), demonstrating the negative impact of anomalies in the model. The temporal autoregression parameter, $\phi$, was slightly overestimated by the model (Fig~\ref{fig:phi}). This was also true of the spatial autoregression parameters; the model underestimated the partial sill, $\sigma^2_{TD}=3$, and overestimated the nugget effect, $\sigma^2_{NUG}=0.1$ and the range parameter, $\alpha_{TD}=10$ (Fig~\ref{fig:spat_pars}).

\section{Case study: analysis and results}
\label{sec:cstudy}

In this section, we delve into the analysis of the motivating dataset introduced in Section 2. 

\subsection{Analysis of water level using IMPALE - A data pre-processing framework}

First, we will implement a pre-processing step to enhance the data quality and improve overall performance of the anomaly detection approaches.
The objective of this process is to discern events that can subsequently serve as covariates in the statistical model, fill in missing covariate data gaps, and evaluate the similarity among sites.

In this study, we introduce the IMPALE framework, a three-step approach designed specifically for water level analysis. 
IMPALE involves the following steps: IMPute and fill the gaps, ALign the time series, and assess the relationship and spatial dependence among sensors, and carry out Events identification.

\begin{figure}[h]
  \centering
   \includegraphics[width=5.5in]{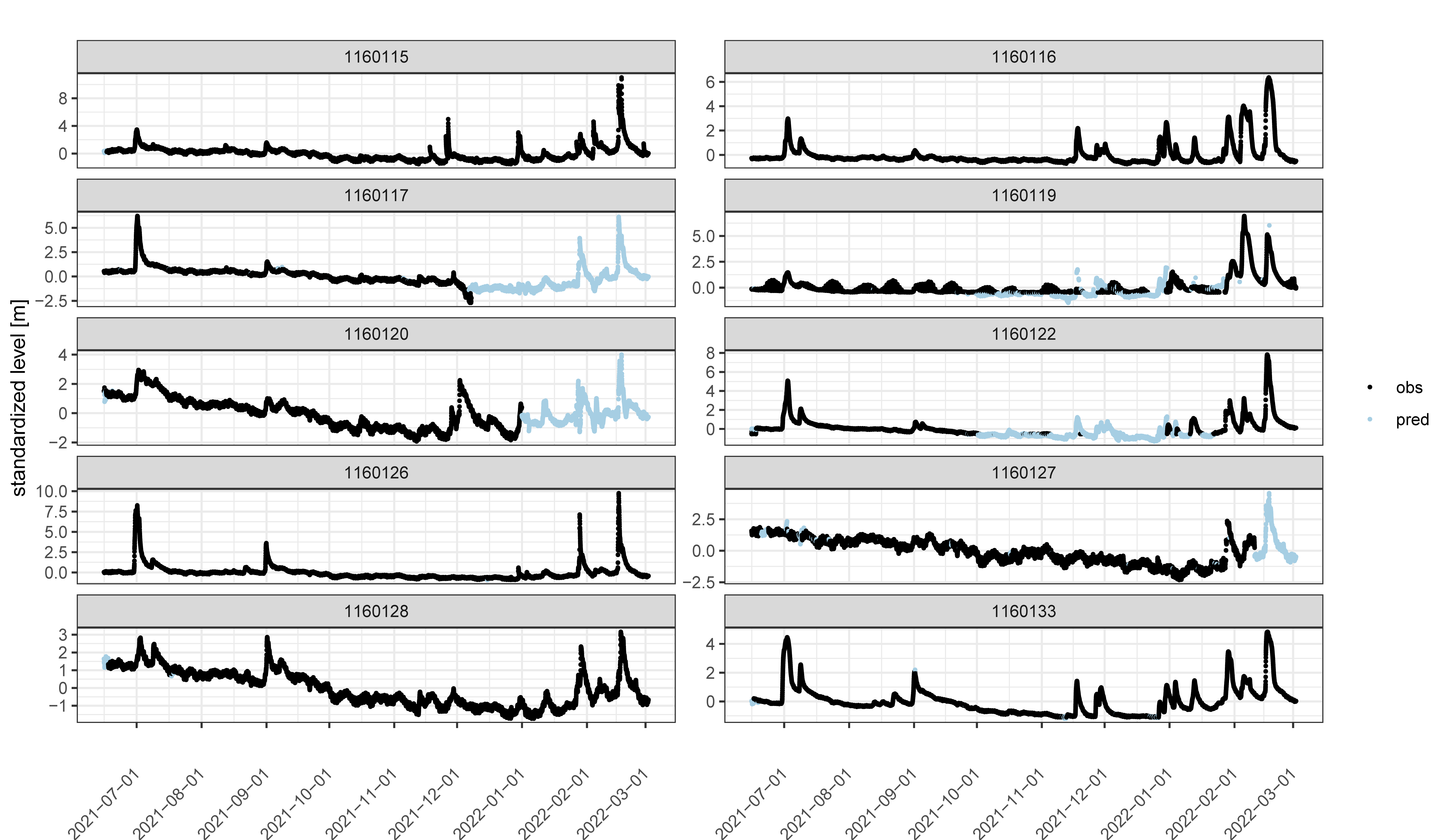}
  \caption{Time series of standardized water level data. Standardization was performed by subtracting the mean and dividing by the standard deviation for each site. Imputed values are represented in light blue. }
    \label{fig:level_imputed}
\end{figure}
\begin{figure}[h]
  \centering
   \includegraphics[width=6.5in]{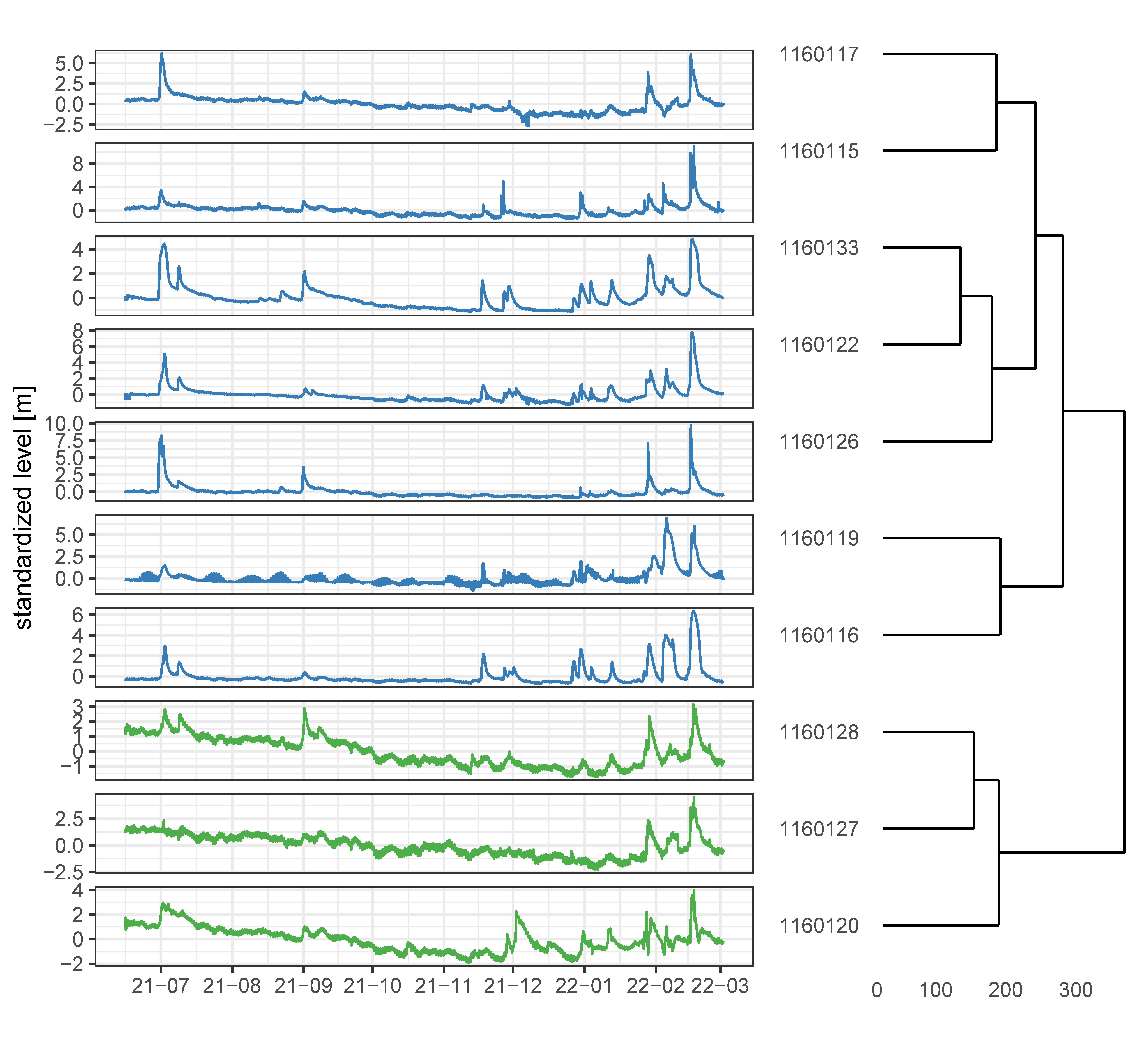}
  \caption{
  Time series of standardized water level data (left) and dendrograms with the clustering structure among sites (right).  }
  \label{fig:level_clusters_dtw}
\end{figure}
\begin{figure}[h]
  \centering
   \includegraphics[width=6.5in]{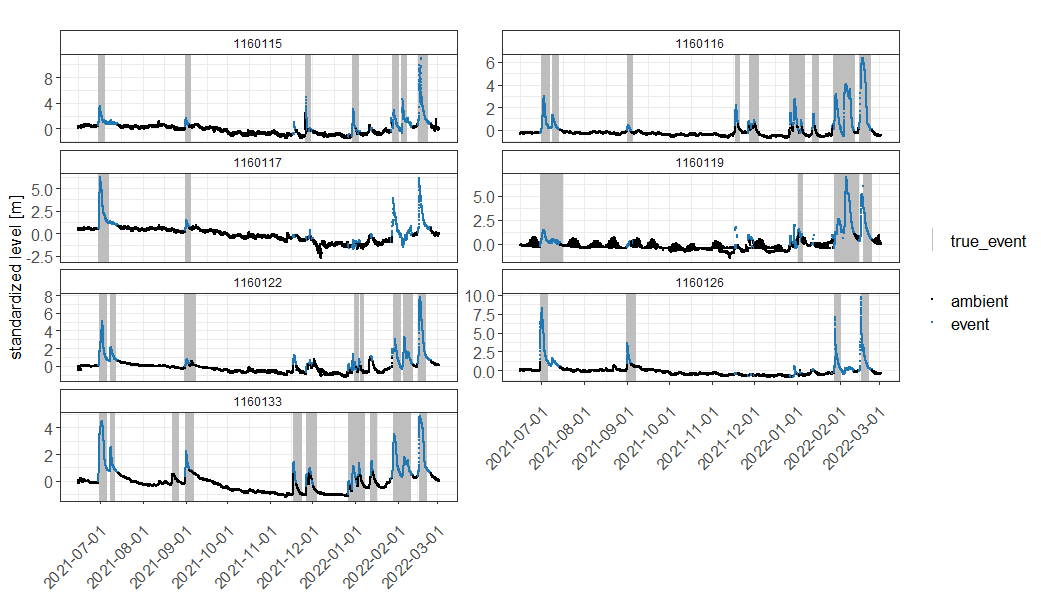}
  \caption{
  Ambient and events identified in the time series of standardized water level data using a Multivariate Hidden Markov Model (MHMM). } 
  \label{fig:mhmm}
\end{figure}

\begin{enumerate}
    \item  {\bf Imputation}: water level sensors often get out of the water producing unreliable level data. 
    Note that this might or might not cause anomalies in the other parameters such as TSS as the sensors might be independent.
    We start by performing multivariate time series imputation using the R package \emph{mtsdi} \cite{mtsdi}. The results from the imputation are shown in Fig~\ref{fig:level_imputed}. An alternative imputation using Bayesian inference can be implemented using package \emph{SSNbayes} \cite{santos2022ssnbayes}, which allows capturing spatial dependence and propagating the uncertainty of the prediction on water level into the TSS estimation. 
    
    \item {\bf Alignment}: optimal match and alignment of the time series was obtained using Dynamic Time Warping (DTW) which minimizes the sum of absolute differences among time series \cite{senin2008dynamic}. This approach allows clustering sites based on similarity or time series distance \cite{aghabozorgi2015time}. Fig~\ref{fig:level_clusters_dtw} shows the clusters formed performing DTW (two clusters in blue and green) and dendrograms with the clustering structure among sites. For the rest of this work and for the sake of simplicity we excluded three sites with very different behaviour and use the seven sites in cluster one. This ensures good spatial association between sites and will help borrow strength from the neighbours when detecting anomalies. In practice, these three sites can be analysed as a separate cluster or individually. Not done here for the sake of space.
    These sites are geo-spatially distinct from the rest, with 1160120 serving as the reference upstream site for the Herbert River, and 1160127/1160128 located along the southern Stone River catchment. Due to the period of observation analyzed, it is probable that a significant portion of the rainfall occurred in the lower catchment, leading to a distinct distribution of rainfall events throughout the time series. Therefore, it is not unexpected to observe these sites clustering separately from the remaining sites.
    

    \item  {\bf Events identification}: Water quality and discharge events tend to produce higher levels of turbidity, nutrients and contaminants. It is critical to identify these water quality events as they help differentiating sensor anomalies (which do not occur across multiple sites at the same periods of time).  We used a Multivariate Hidden Markov Model \cite{zucchini2009hidden} to estimate the latent state of the catchment. This approach produces ``ambient'' and ``event'' labels for the catchment for each time points. 
    We show in Fig~\ref{fig:mhmm} the separation of ambient (black) and event (blue). The vertical gray lines are the true water events produced by water quality specialists assumed the ground truth.   
    Note that on site 1160117, no ground truth was produced during 2022, as no data was available. However, we show the labels for the imputed data from the MHMM. This method was also effective in a site like 1160119 that is affected tidal influence.
    Overall, we found that the MHMM is effective in the identification of events. The overall accuracy was 87.89\%. 
    In total, 80\% of the data points pertaining to events where correctly classified (sensitivity) and the specificity was 89.6\% with a MCC = 62.62. 
    The Brier score, as defined by \cite{brier1950verification}, was calculated as $\frac{1}{N} \sum_{i=1}^{N} (f_i - o_i) ^ 2$, where $f_i$ represents the predicted and $o_i$ is the observed outcome.
    In this MHMM analysis, the Brier score was 0.121. 
   
\end{enumerate}

\subsection{Analysis of total suspended solids (TSS)}

{\bf Anomaly detection using recursive learning}

We developed a recursive online method for fitting vector autoregressive Bayesian spatio-temporal regression models. Details of the general Bayesian recursive model can be found in \citeA{sarkka2013bayesian}.

We denote the total suspended solids as $TSSeq$, which is modelled as a function of the water level ($level$), a binary variable indicating event vs ambient conditions obtained in previous section ($event$) and precipitation data ($prec$). This can be represented more formally using mathematical notation as:

\begin{equation}
TSSeq = \beta_0 + \beta_1 \cdot level + \beta_2 \cdot event + \beta_3 \cdot prec  
    \label{eq:lmcs}
\end{equation}

\noindent where $\beta_0$ is an intercept and $\beta_1, \beta_2,$ and $\beta_3$ are the coefficients associated with each predictor variable. Note that Equation (\ref{eq:lmcs}) corresponds to $\bm{X}_{t}\bm{\beta}$ in Equation (\ref{eq:err_}).

The data has a temporal resolution of one observation every 30 minutes. For ease of processing, we organized the data into monthly batches. This means that we took one month of data, fitted the model, estimated the parameters, and identified anomalies.
In each batch of data, we derive posterior predictive distributions for the spatial and temporal parameters ($\sigma_{td}$, $\alpha_{td}$, and $\phi_s$), along with the regression coefficients ($\beta$). The posterior distributions obtained from each batch of data then served as priors for subsequent batches, allowing for the recursive updating of the model's parameters.

Batches with anomalous behaviour did not inform the estimation of parameters. Specifically, if the i-th batch contained anomalous data, the prior distributions for the subsequent batch were formed from the last non-anomalous batch. This approach ensures that the estimated parameters are less influenced by anomalous data and reduces the bias that may arise from such data.

For more effective online updating and to account for the dynamic change in the environment, models can be fitted to smaller chunks of data, such as hourly, daily, and weekly intervals.
The computational time for fitting these models for all batches of data was smaller than the frequency of the data ($<$ 30 mins).

We employed the PPD approach, which demonstrated suitable performance in our simulation study and does not require fitting additional methods such as mixtures and HMMs. Although we only utilized one iteration to avoid recomputing the models, we acknowledge that using two iterations may lead to further improvements in performance. However, we maintain that our chosen approach provides a suitable level of accuracy and practicality for the purposes of this study.

\begin{figure}[h]
  \centering
   \includegraphics[width=6.5in]{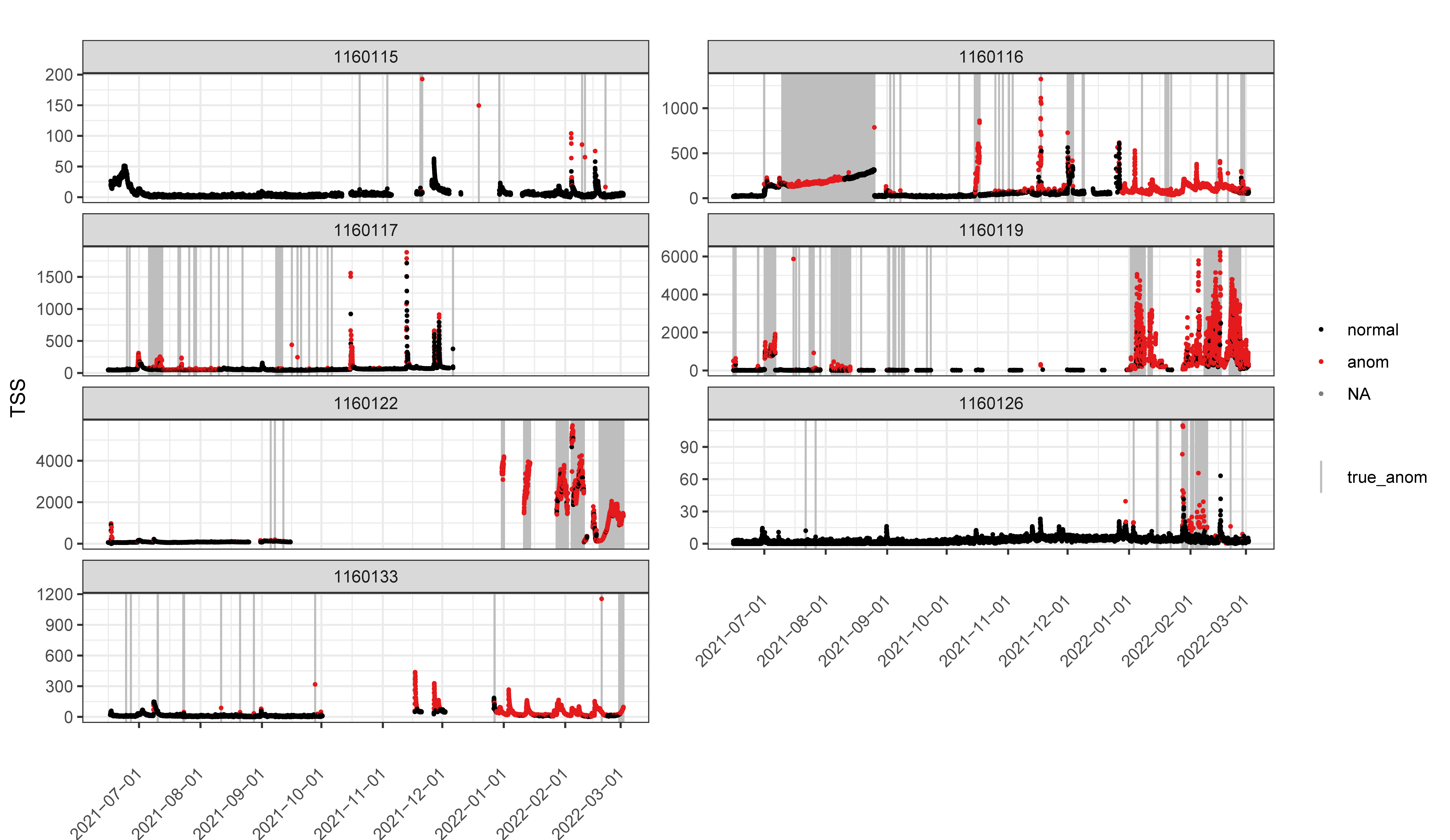}
  \caption{
  Anomalies identified in the turbidity (TSS) data.  
 The gray vertical lines and shaded areas delineate labeled anomalies (ground truth).  Points in red represent the predicted anomalies from the models.
  }
  \label{fig:tur_anom_vs_labels}
\end{figure}

For each observation, we computed the posterior predictive distribution, and identified any observation falling outside the 95\% highest density interval (HDI) as an anomaly. 
The identified anomalies were highlighted in red in Fig~\ref{fig:tur_anom_vs_labels}, with the (true) anomalous sections in the data represented with gray vertical lines. 
As previously mentioned, the HDI is a credible interval used to measure the uncertainty of the posterior distribution and the likelihood of observing new data. 
The 95\% HDI, in particular, represents the range of values that contain 95\% of the probability mass for the posterior distribution.

Our model demonstrates suitable performance in detecting drift (in site 1160116), sudden shifts, and periods of high variability (in 1160119). Additionally, it accurately identified the constant offset present in 1160122.
Overall, using this approach, we were able to identify and highlight the anomalous observations with high accuracy.

The table \ref{table:perform} presents the performance measures of a spatio-temporal anomaly detection method for the seven types of anomalies in TSS data. The performance measures include sensitivity ($se$), specificity ($sp$), accuracy ($acc$), adjusted accuracy ($acc_{adj}$), Matthews correlation coefficient ($MCC$), and Brier score.

The results show that the PPD method performs well in detecting most types of anomalies. 
Across all anomaly types, 74\% of anomalous observations (in the high-frequency data) were correctly identified and the accuracy and adjusted accuracy were 83.6 and 79.2\%. 
Note that the overall $sp=84\%$ (not reported for individual types of anomaly as they are all the same).

Analysing by type of anomaly, the highest sensitivity is observed for the constant offset ($se=90\%$). 
The highest overall accuracy and adjusted accuracy are observed for the constant offset type of anomaly, while the highest MCC is observed for the sensor drift. 
The Brier score, which measures the accuracy of the method's probability estimates, is low for all types of anomalies, indicating that the PPD method provides reliable probability estimates. 
The results in the table indicate that the PPD method is effective in detecting various types of anomalies in TSS data, making it a valuable tool for online monitoring of water quality in spatio-temporal stream networks.

We conducted similar analyses on the same catchment using TSS data obtained from low-cost sensors, which posed greater challenges for the identification of anomalies. The results of these analyses are presented in \ref{sec:appB}.

\begin{table}[h]
\centering
\caption{Performance measures in the detection of anomalies in turbidity data by type of anomaly in the labelled dataset and overall.}
\label{table:perform}
\begin{tabular}{llrrrrrr}
  \hline
Anomaly type & method & se & sp & acc & acc\_adj & MCC & Brier \\ 
  \hline
Small spike & ppd & 0.4530 &  & 0.8453 & 0.6496 & 0.0381 & 0.1547 \\ 
Large spike & ppd & 0.6774 &  & 0.8459 & 0.7618 & 0.0595 & 0.1541 \\ 
Cluster of spikes & ppd & 0.8276 &  & 0.8461 & 0.8369 & 0.0428 & 0.1539 \\
Sudden shift & ppd & 0.5940 &  & 0.8443 & 0.7201 & 0.1024 & 0.1557 \\ 
High variability & ppd & 0.7965 &  & 0.8451 & 0.8213 & 0.2474 & 0.1549 \\ 
Constant offset & ppd & 0.9017 &  & 0.8469 & 0.8739 & 0.2361 & 0.1531 \\ 
Sensor drift & ppd & 0.7074 &  & 0.8389 & 0.7768 & 0.3186 & 0.1611 \\ \hline
  All & ppd & 0.7397 & 0.8461 & 0.8363 & 0.7929 & 0.4187 & 0.1637 \\
   \hline
\end{tabular}
\end{table}

\section{Discussion and conclusions}
\label{sec:Dis}

Water quality data from in-situ sensors can be challenging to analyse due to the presence of technical anomalies as well as water quality events and spatio-temporal dependencies. Here, we have introduced and successfully demonstrated a general suite of unsupervised statistical methods to detect technical anomalies in data arrays from in-situ sensors used for water-quality monitoring in stream networks.

Our results show that these tools are suitable for detecting challenging technical anomalies in high-frequency data produced by sensor technology. 
The four approaches discussed were more effective in the identification a range of anomalies, compared to the ARIMA, which can only describe temporal correlation in the data at individual sites. These spatially aware models capture events happening simultaneously across multiple locations, allow more effective data imputation of time series and help identify spatial anomalies.
Since these models were set within a Bayesian framework, we can obtain probabilistic estimates, anomalous scores and quantify the uncertainty in the data.

The findings of this study were reinforced through a comprehensive simulation study that incorporated various datasets and types of anomalies. Nonetheless, future investigations could take into account scenarios where the model assumptions are violated, such as deviations in the temporal structure and non-linear associations between the response variable and covariates. These circumstances can help to provide a more complete understanding of the model's behavior and its robustness under different scenarios of data generation. Therefore, additional research in these areas could provide valuable insights and enhance the overall applicability of the proposed methodology.


We chose to implement a spatio-temporal stream-network model \cite{santos2022bayesian} because the unique characteristics of streams (e.g. branching network topology, flow connectivity and directionality), together with high-frequency sampling tended to produce data with spatial and temporal dependencies. However, analysing the empirical residuals from the model fit via finite mixtures and HMMs offers a flexible set of solutions to distinguish anomalous from normal observations, regardless of the initial approach used. For example, the spatio-temporal model could be replaced with other approaches that produce predictions, such as Gaussian additive models (GAMS) \cite{wood2017generalized}, time series models \cite{hyndman2018forecasting}, or machine learning algorithms \cite{chandola2009anomaly}.

Physical models are constructed based on well-established physical principles that govern processes within river networks. These models are capable of simulating water quality parameters across diverse environmental conditions, offering a dynamic representation of system behavior. 
By comparing observed data with model predictions, physical models can detect discrepancies, potentially identifying anomalies in the dataset. 
An extensive body of literature has introduced various physical methods for identifying anomalous data and change points in time series analysis \cite{zegre2010lieu, housh2017integrating}.
Since these models can capture intricate interactions and dependencies within the river system, they can complement the capabilities of statistical methods. 
However, these physical models can be very difficult to develop, require many parameters (which are often unidentifiable from available data) and the models can very computationally expensive to form predictions.
Future research endeavors could explore the development of efficient hybrid approaches, merging physical modeling with Bayesian statistical methods, to achieve enhanced performance and gain deeper insights into water quality dynamics.

Water quality parameters often are non-normally distributed and can be affected by censorship or zero-inflation, posing a challenge for statistical analysis. To address this issue, current practice involves using log transformations to normalize the data. Alternatively, our modeling framework can be adapted to follow the Generalized Linear Model (GLM) philosophy, which provides a flexible approach to analyzing non-normal data and can account for the various distributional characteristics of water quality parameters.

From a practical point of view, using methods such as mixtures and HMMs involves challenging issues such as the selection of the number of components.
  This, however, could be approached using common model-selection measures such as Akaike information criterion (AIC), Bayesian information criterion (BIC) and widely applicable information criterion (WAIC) to compare competing models. Other natural extensions of these methods include multivariate approaches to explain the complex relationships between multiple correlated parameters such as nitrate, turbidity, and conductivity \cite{kermorvant2021understanding}. 
The tools can also be easily extended to build statistical process control approaches (e.g. multivariate cumulative sum (MCUSUM) or multivariate exponentially weighted moving average (MEWMA) control charts \cite{montgomery2020introduction, santos2013MSQC}. Similarly, the PPD Bayesian method can be extended by constructing multivariate credible intervals \cite{besag1995bayesian}. We are currently exploring multiple variations of these models.

The motivation for developing these methods was primarily for automating the Quality Assurance and Control (QA/QC) processes for high-frequency in-situ data, but they could also be applied retrospectively. For example, anomalies could be identified and removed during the modelling phase to obtain less biased estimates of the parameters of interest (e.g. regression coefficients). This is important in cases where data are released for public use, especially where the presence of anomalies due to sensor issues could reduce confidence in the quality of the data.

This work opens up new avenues of research, especially around the detection of anomalies in space-time data in streams. The outcomes of this research are paramount for the delivery and presentation of environmental near real-time data to the public. As more high-frequency monitoring data are made available to the general community, timely detection of the anomalies/errors will be imperative for gaining and maintaining trust and transparency in the measured data. This is particularly important when data are used to identify sources of water quality pollutants and demonstrate to landholders the benefits associated with changes in land management practices. Future research should explore the integration of the statistical models here described with other statistical quality control tools for high-frequency data produced by arrays of sensors.

\section{Open Research}

\noindent Software: The spatio-temporal stream network models were fit using the R package \textsf{SSNbayes} \cite{santos2022ssnbayes} [Software].
R code used to simulate spatio-temporal stream network data and to incorporate anomalies can be found at \cite{edgar_santos_fernandez_2023}. \\
Data: The Herbert river data [Dataset] from the case study is attached to this submission and will be deposited in this repository upon acceptance.






\acknowledgments

This research was supported by the Australian Research Council (ARC) Linkage Project LP180101151: ``Revolutionising water-quality monitoring in the information age''.
We thank the Water Quality \& Investigations group from the Queensland Department of Environment and Science for the installation of the water quality sensors and for producing the labelled dataset used in the case study, specially Matthew Sinclair and Ben Ferguson. 
We also thank Puwasala Gamakumara for developing the R script for accessing the in-situ data from low-cost sensors.
The authors are grateful to the six reviewers and Associate Editor for their insightful comments.

\bibliography{ref.bib}

\clearpage

\appendix

\section{Other results}

\begin{figure}[htbp]
  \centering
   \includegraphics[width=7in]{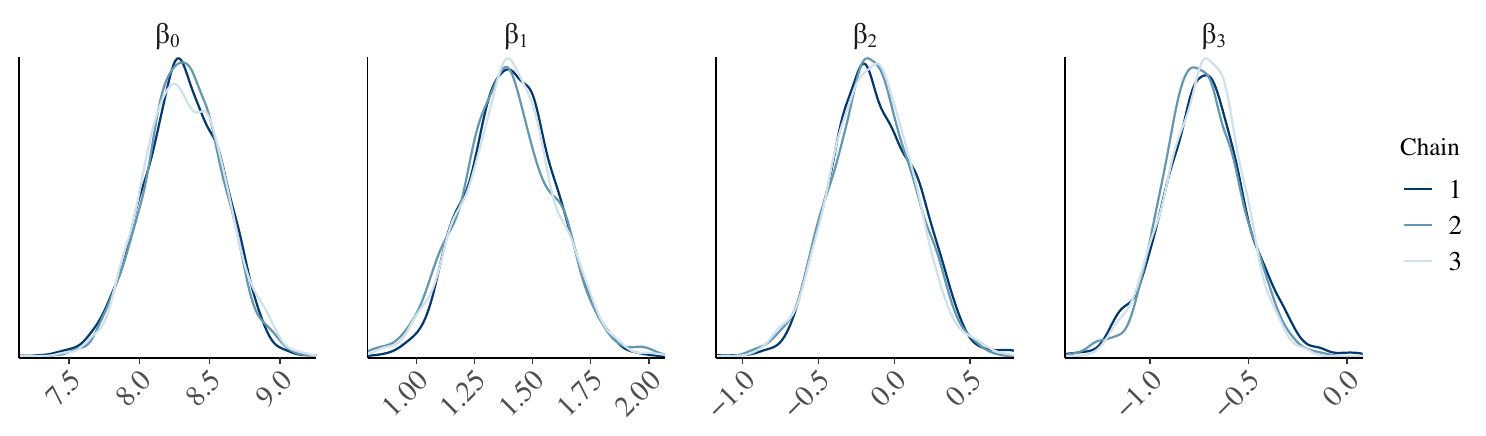}
  \caption{Posterior distribution of the regression coefficients. True values $\beta_0 = 10$, $\beta_1 = 1$, $\beta_2 = 0$ and $\beta_3 = -1$. }
  \label{fig:betas}
\end{figure}

\begin{figure}[htbp]
  \centering
   \includegraphics[width=5in]{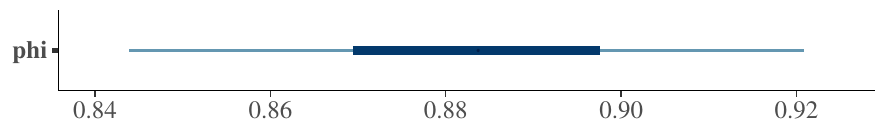}
  \caption{Posterior boxplot of the autoregression parameter. The true value is $\phi = 0.80$.  }
  \label{fig:phi}
\end{figure}

\begin{figure}[htbp]
  \centering
   \includegraphics[width=7in]{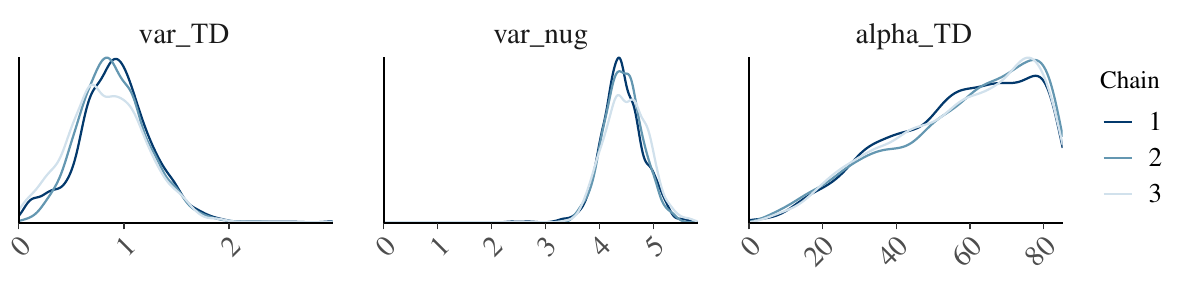}
  \caption{Posterior distribution of the spatial parameters. True values: $\sigma^2_{TD}=3$, $\alpha_{TD}=10$ , $\sigma^2_{NUG}=0.1$ }
  \label{fig:spat_pars}
\end{figure}

\clearpage

\section{Extended analysis using low-cost sensors}
\label{sec:appB}

\begin{figure}[htbp]
  \centering
   \includegraphics[width=5in]{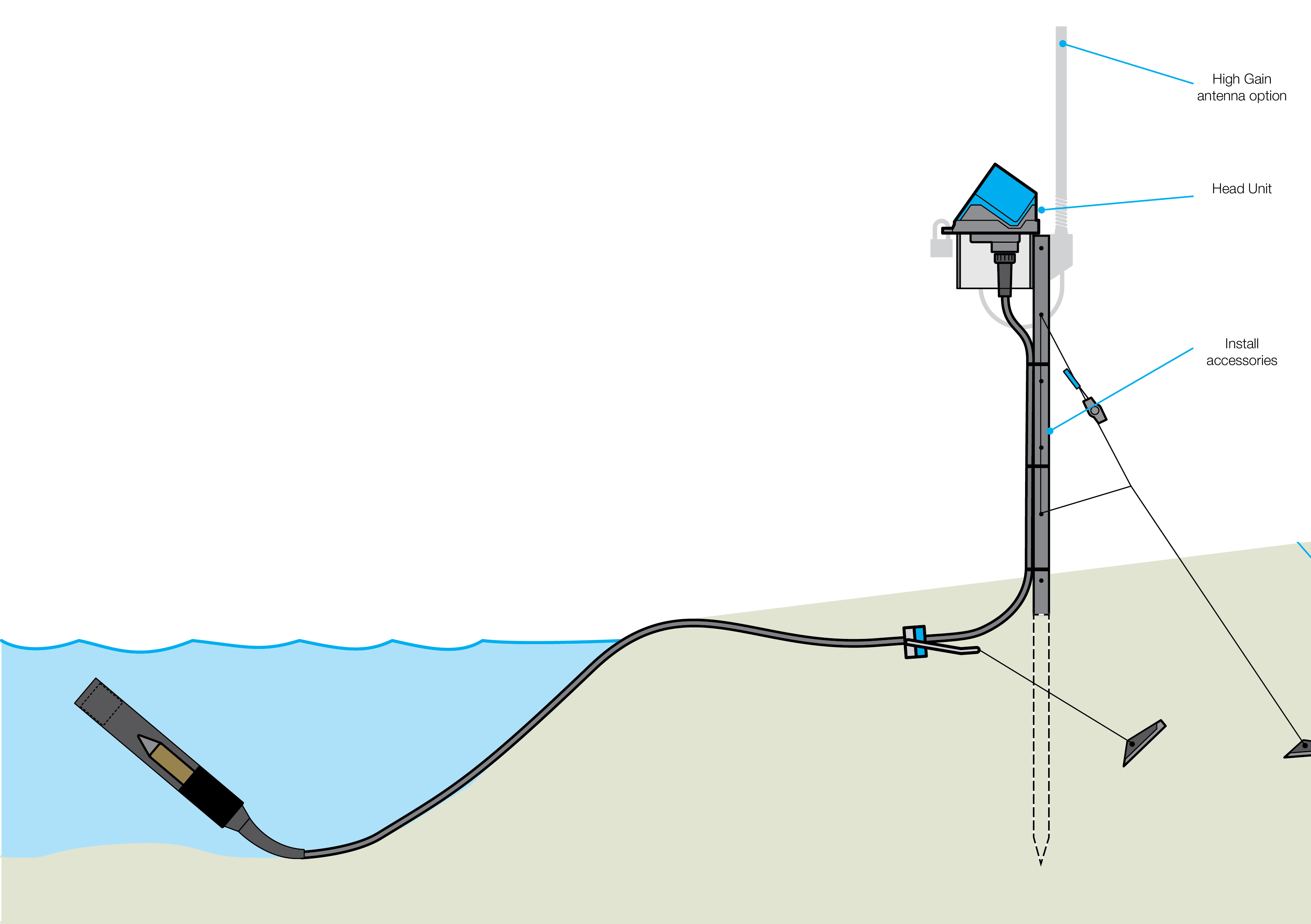}
  \caption{Diagram of the BlueSpots used in the Herbert River. }
    \label{fig:bluespots}
\end{figure}

In this section we extend the analyses from Section \ref{sec:cstudy} using data from low-cost sensors. 
We installed five low-cost BlueSpot water-quality sensors (model TLS02 021; Fig~\ref{fig:bluespots}) produced by IntelliDesign in the branching network of the Herbert River. These Internet of Things (IoT) devices measure stream water level (m), specific conductivity ($\mu$S/cm) and turbidity (NTU). Measurements were recorded every 15 minutes and uploaded via the 4G mobile network to a database. Each BlueSpot was installed at the edge of the riverbank in a location best representing that part of the waterway. The BlueSpot telemetry unit was attached to a star picket anchored to the ground using cables. The probe itself was suspended in the water column approximately 30cm above the riverbed or riverbank depending on where it was installed. 
We used time-series data for turbidity and water level collected between 16-Feb-2021 and 04-Mar-2021 in this study.

\subsection{Results}

We used the best-performing methods from the simulation study (PPD and HMM ) to identify anomalies in water level and turbidity measurements. 
We first fit a space-time model introduced in Equation (\ref{eq:eq1}) to identify anomalies in water level using a temporal and two spatial covariates:

\begin{itemize}
    \item precipitation (in mm) obtained from a monitoring station in the Herbert River. \url{https://water-monitoring.information.qld.gov.au/};  
    \item proportion of stream with land uses where fertilizer is likely to be used; and
    \item proportion of natural forest cover in the catchment.
\end{itemize}
The spatial covariates were sourced from the Australian National Environmental Stream Attributes dataset \cite{stein2012national}.

Next we fit a spatio-temporal stream-network model for (log) turbidity using the same three covariates described above, plus the (log) water level. All the methods were then used to predict anomalies. 
 
\begin{figure}[hp]
  \centering
   \includegraphics[width=6.5in]{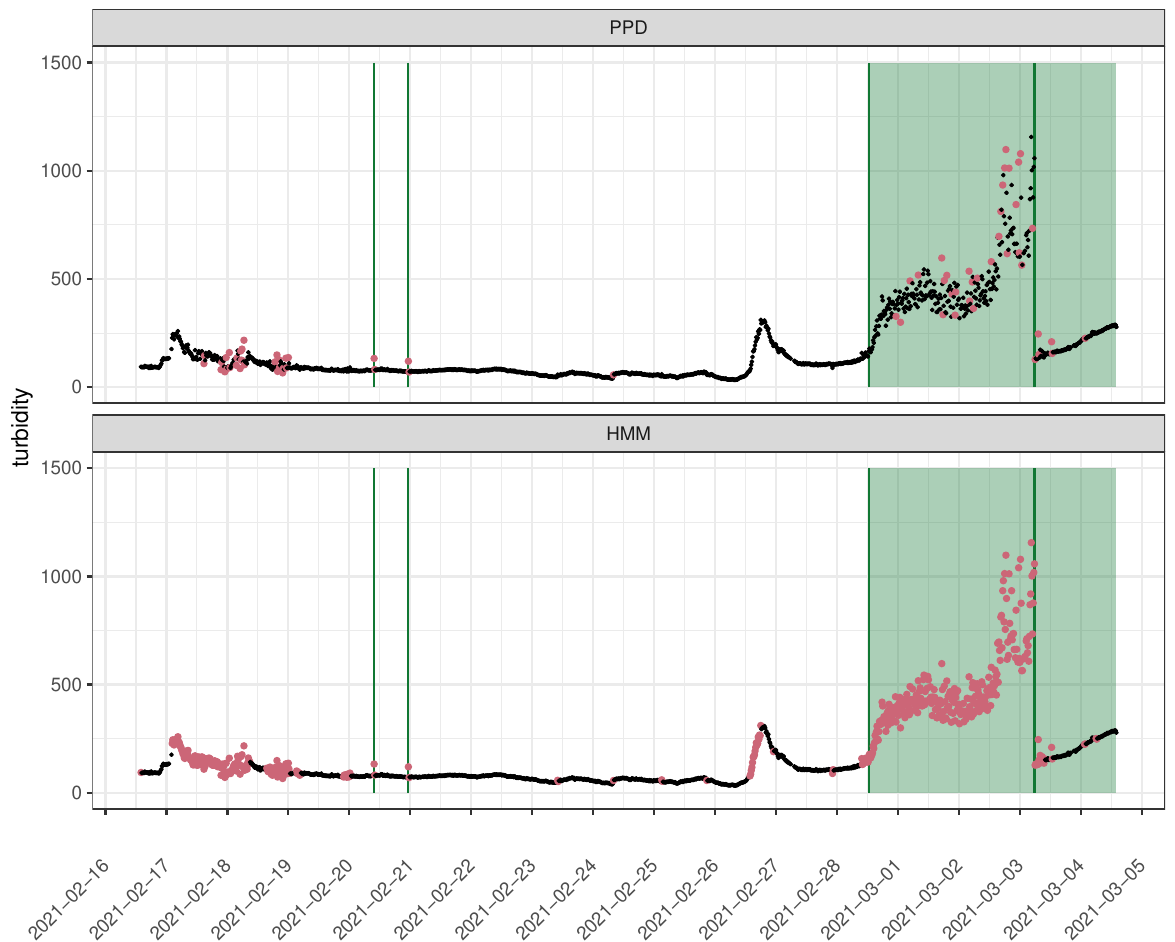}
  \caption{
 Time series of turbidity data and anomalies from Bluespot sensor 1 using (a) PPD and (b)HMM. 
 The green vertical lines and shaded areas delineate labeled anomalies.  Points in red represent the predicted anomalies from the models.}
  \label{fig:turbPPD}
\end{figure}

For illustration, let us consider the turbidity from Bluespot sensor 1, where some anomalous behavior occurred (Fig~\ref{fig:turbPPD}). 
In green, we represent the {\it data points (with vertical lines)} and {\it periods (with rectangles)} where true technical (sensor) anomalies are present in the labelled dataset. 
In red we represented the predicted anomalies using the PPD method. 
Notice the two sudden spikes on the 20-Feb-2021 at 19:45 and 21-Feb-2021 at 09:15. There is also a period of high variability (28-Feb-2021 to 03-Mar-2021) followed by an ongoing drift (3-Mar-2021 16:00) (Fig~\ref{fig:turbPPD}).

The two approaches successfully identified the four anomalies in the data, including the start of the drift event.
The PPD model identified the high variability event, in addition to the two sudden spikes and the beginning of the drift (Fig~\ref{fig:turbPPD}a). 

However, the HMM method was more sensitive to changes in the process than the other two models (Fig~\ref{fig:turbPPD}c). Interestingly, the HMM models also identified a true water quality event around 26-Feb-2021, which was observed across the other Bluespot sensors.

\end{document}